\documentclass[twocolumn]{aastex62}

\hypersetup{linkcolor=blue,citecolor=blue,filecolor=cyan,urlcolor=blue}

\usepackage{braket}
\usepackage{amsmath}
\usepackage{bm}
\usepackage{mathrsfs}

\newcommand{\lya}{Ly$\alpha\,$}
\newcommand{\lyb}{Ly$\beta\,$}

\newcommand{\heii}{\ion{He}{2}}
\newcommand{\cii}{[\ion{C}{2}]}

\newcommand{\mgii}{\ion{Mg}{2}}

\newcommand{\oiii}{[\ion{O}{3}]}

\defcitealias{Eilers2020}{Paper I}
\revised{\today}
\submitjournal{ApJ}

\shorttitle{Four Quasars at $z\sim 6$ with Lifetimes $<10^4$~years}
\shortauthors{A.-C. Eilers et al.}

\begin{document}\frenchspacing

\title{\textbf{Detecting and Characterizing Young Quasars II: Four Quasars at $z\sim 6$ with Lifetimes $<10^4$~years}}

\author[0000-0003-2895-6218]{Anna-Christina Eilers}\thanks{NASA Hubble Fellow}
\affiliation{MIT Kavli Institute for Astrophysics and Space Research, 77 Massachusetts Ave., Cambridge, MA 02139, USA}

\author[0000-0002-7054-4332]{Joseph F. Hennawi}
\affiliation{Physics Department, University of California, Santa Barbara, CA 93106-9530, USA}

\author[0000-0003-0821-3644]{Frederick B. Davies}
\affiliation{Lawrence Berkeley National Laboratory, CA 94720-8139, USA}
\affiliation{Max-Planck-Institute for Astronomy, K\"onigstuhl 17, 69117 Heidelberg, Germany}

\author[0000-0003-3769-9559]{Robert A. Simcoe}
\affiliation{MIT Kavli Institute for Astrophysics and Space Research, 77 Massachusetts Ave., Cambridge, MA 02139, USA}

\correspondingauthor{Anna-Christina Eilers}
\email{eilers@mit.edu}

\begin{abstract}
 The extents of proximity zones of high--redshift quasars enable constraints on the timescales of quasar activity, which are fundamental for understanding the growth of the supermassive black holes (SMBHs) that power the quasars' emission. 
 In this study, we obtain precise estimates for the ultraviolet (UV) luminous lifetimes of ten quasars at $5.8< z< 6.5$. These objects were pre--selected to have short lifetimes based on preliminary measurements of their proximity zone sizes, and were then targeted for high quality follow--up sub--mm, optical, and infrared observations required to increase the measurements' precision and securely determine their lifetimes. 
 By comparing these proximity zone sizes to mock quasar spectra generated from radiative transfer simulations at a range of different lifetimes, we deduce extremely short lifetimes $t_{\rm Q}<10^4$~yr for four objects in our sample, whereas the remaining quasars are consistent with longer lifetimes of $t_{\rm Q}\gtrsim 10^5$~yr. These young objects with small proximity zones represent $\lesssim10\%$ of the quasar population as a whole. 
 We compare our results in detail to other studies on timescales of quasar activity, which point towards an \textit{average} lifetime of $t_{\rm Q}\sim10^6$~yr for the quasar population. This is consistent with finding newly turned on quasars approximately $\sim 1-10\%$ of the time. These young quasars represent an unique opportunity to study triggering and feedback mechanisms of SMBHs, since the onset of their UV luminous quasar phase happened only recently, and therefore traces of this process might still be observable. 
\end{abstract}


\section{Introduction}\label{sec:intro}

Since the discovery of the first high--redshift quasars at $z\gtrsim 6$ nearly two decades ago \citep{Fan2000, Fan2001}, these distant luminous objects have been employed extensively as beacons through the cosmic web illuminating the diffuse intergalactic medium (IGM) and providing unique insights into the young universe. 
The evolution of the Lyman series absorption features observed in the quasars' spectra are one of the key observational probes of the physical properties of the intergalactic gas, which includes the majority of baryons in the universe. 

The applications of these absorption features are abundant: for instance, the evolution of the Lyman-$\alpha$ (\lya) and Lyman-$\beta$ (\lyb) opacities at high--redshifts have provided invaluable constraints on the timing and morphology of the Epoch of Reionization \citep[e.g.][]{Fan2006, Becker2015, Eilers2018a, Eilers2019b, Bosman2018, Keating2020, Yang2020b}. 
Statistical properties of the absorption lines in the Lyman-series forest such as the probability distribution function \citep[e.g.][]{Lee2015, Cieplak2017, Eilers2017b, Davies2018a} or the power spectrum \citep[e.g.][]{Palanque2013, Irsic2017a, Walther2019, Boera2019} encode information about the physical properties of the absorbing gas, e.g.\ its thermal state and the ionizing UV background, whereas metal absorption lines in the unabsorbed part of the quasar spectra provide valuable clues about the chemical enrichment of the IGM \citep[e.g.][]{Simcoe2012, Dodorico2013, Cooper2019, Becker2019, Simcoe2020}. 

However, while the aforementioned studies exploit the quasar spectra to constrain the properties of the \textit{intergalactic gas}, we can in turn also explore the characteristics of the \textit{quasars} by means of the IGM absorption features, since the quasars' ionizing radiation illuminates the surrounding gas in their vicinity, injects heat and changes its ionization state. Thus by studying the properties of the IGM close to quasars, we can set constraints on the duration and intensity of the emitted ionizing radiation, as well as the obscuration fraction of the accreting black hole and the radiation geometry \citep[e.g.][]{Khrykin2016, Khrykin2019, Eilers2017a, Eilers2018b, Davies2019b, Davies2019a, Schmidt2017, Schmidt2018, Bosman2019}. 

In this paper we will focus on constraining the timescales of the nuclear activity of quasars, i.e. the time during which the central supermassive black holes (SMBHs) actively accrete material from their environments, powering the quasar's UV luminous emission. We will distinguish between different timescales of quasar activity as follows: 
We define the \textit{lifetime} of quasars $t_{\rm Q}$ (or equivalently their \textit{age}) as the time at which the current UV luminous quasar phase began, i.e. the quasar turned on at a time $t_0-t_{\rm Q}$ in the past. 
In this case, $t_0$ denotes the time at which the radiation observed today was emitted, which corresponds to the quasar's emission a lookback time ago. 
However, quasar activity may be episodic with alternating epochs of luminous quasar and quiescent phases. The quasar's \textit{duty cycle} indicates the integrated time over the age of the Universe at that epoch that galaxies shine as active quasars, and thus provides an upper limit on the lifetime. 
A quasar's lifetime and duty cycle are equivalent in the case of a simple ``light--bulb'' light curve model, in which the quasar is assumed to emit at constant luminosity for its entire lifetime. 

These timescales of quasar activity represent important parameters for understanding the growth of their accreting SMBHs, which can have masses up to $\gtrsim10^9\,M_{\odot}$ as early as $<1$~Gyr after the Big Bang \citep[e.g.][]{Mortlock2011, Mazzucchelli2017, Banados2018, Yang2020a, Wang2021}. Assuming a constant supply of fueling material, black holes grow exponentially during the lifetime of the quasar from an initial black hole seed with a mass $M_{\rm seed}$, i.e. 
\begin{equation}
    M_{\rm BH}(t_{\rm Q})=M_{\rm seed}\cdot \exp\left(\frac{t_{\rm Q}}{t_{\rm S}}\right).  \label{eq:mbh}
\end{equation}
The growth of the SMBHs occurs on a characteristic time scale, which is called the ``Salpeter'' time $t_{\rm S}$ \citep{Salpeter1964} or e-folding time, i.e. 
\begin{equation}
    t_{\rm S} = 0.45 \left(\frac{\epsilon}{1-\epsilon}\right)\left(\frac{L_{\rm bol}}{L_{\rm Edd}}\right)^{-1}~\rm Gyr, \label{eq:salpeter}
\end{equation}
where $\epsilon$ denotes the radiative efficiency of the accretion with a fiducial value of $\sim10\%$ for thin accretion disks \citep{ShakuraSunyaev1973}, and $L_{\rm bol}$ describes the bolometric luminosity of the quasar, which has a theoretical upper limit of the Eddington luminosity $L_{\rm Edd}$. In order to grow a SMBH with a mass of $\sim10^9\,M_{\odot}$ from an initial black hole seed from a stellar remnant of a population III star, i.e. $M_{\rm seed}\sim 100\,M_{\odot}$ \citep[e.g.][]{MadauRees2001}, it requires at least $16$ e-foldings, i.e. $t_{\rm Q}\gtrsim 7 \times10^8$~yr, even if the quasars accrete continuously at the Eddington limit \citep[e.g.][]{Volonteri2010, Volonteri2012}. However, it is currently unknown whether quasars obey this exponential light curve, or if other physics related to the triggering of quasar activity and the supply of fuelling material complicate this simple picture, which would give rise to more complex ``flickering'' light curves \citep[e.g.][]{DiMatteo2005, Springel2005, Hopkins2005, Novak2011, Schawinski2015}.

Quasar lifetimes can be estimated by various different methods. Properties of the IGM, such as the opacity, ionization state, and temperature in the vicinity of a quasar react to any changes in a quasar's accretion rate with a time lag. Such an ionization ``echo'' leaves an imprint on the surrounding IGM, which can be observed for instance in the \lya\ opacity of a background source at small impact parameters \citep{Adelberger2004, Hennawi2006, VisbalCroft2008, Schmidt2019}.
Similarly, the time lag between the onset of the quasar's radiation and its arrival at nearby \lya--emitting galaxies (LAEs) can be utilized to constrain the lifetime of  quasars \citep{TrainorSteidel2013, Bosman2019}. The ratio of the number density of dark matter halos hosting an active quasar and the full numbers or halos that could host halos inferred from quasar clustering studies allows for constraints on the quasars' duty cycle \citep[e.g.][]{MartiniWeinberg2001, HaimanHui2001, Shen2007, White2008, ConroyWhite2013}. 

In this study we estimate the lifetime of high--redshift quasars by means of the extents of their \lya\ proximity zones observed in their rest-frame UV spectra along the line-of-sight to the quasars. Proximity zones are the regions of enhanced transmitted flux close to the quasars \citep{Bajtlik1988}, which are sensitive to the lifetime of quasars because the intergalactic gas has a finite response time to the quasars' radiation \citep{BoltonHaehnelt2007a, Khrykin2016, Eilers2017a, Davies2019a, Davies2020, Ishimoto2020}. 
By applying this method to a data set of $31$ $z\sim 6$ quasars we previously discovered a population of very young quasars, which imply extremely short lifetimes of $t_{\rm Q}\lesssim 10^4-10^5$~yr \citep{Eilers2017a, Eilers2018b}. These young objects pose significant challenges for the growth of SMBHs at high redshift, since their lifetimes are several orders of magnitude shorter than the time required for the exponential growth of the black holes described in Eqn.~\ref{eq:mbh} and \ref{eq:salpeter}. 

Here we estimate the lifetime of quasars for ten high--redshift objects, which were pre--selected to likely have
very short lifetimes, i.e. $t_{\rm Q}\lesssim 10^5$~yr, in order to establish a statistically significant sample of known young quasars as well as to estimate the fraction of young quasars within the quasar population at large. To this end, in \citet[][hereafter \citetalias{Eilers2020}]{Eilers2020} we have obtained a multi--wavelength data set for this quasar sample to precisely measure the extents of their proximity zones, which is summarized in \S~\ref{sec:sample}. By comparing these measurements to simulated quasar spectra at different lifetimes from 1D radiative transfer (RT) simulations (\S~\ref{sec:RT}) we will obtain precise lifetime estimates (\S~\ref{sec:lifetimes}). We will compare our results to other studies of quasar activity timescales from the literature (\S~\ref{sec:comparison}) and discuss the implications of very short quasar lifetimes (\S~\ref{sec:discussion}), before summarizing our findings (\S~\ref{sec:summary}). 

Throughout this paper, we assume a flat $\Lambda$CDM cosmology of $h = 0.685$, $\Omega_m = 0.3$, and $\Omega_{\Lambda} = 0.7$.

\section{Quasar Sample and Proximity Zone Measurements}\label{sec:sample}

Our data sample includes ten quasars between $5.8\lesssim z\lesssim 6.5$, which were selected from a parent sample of $122$ quasar spectra. From this sample we selected the best young quasar ``candidates'' whose preliminary measurements for the sizes of the proximity zones $R_{\mathrm p}$ were very small, i.e. $\lesssim 2$~proper Mpc (pMpc) \citepalias[see][for details]{Eilers2020}, and thus they were likely to indicate very short lifetimes. We excluded any quasars showing broad absorption line features (BALs) or proximate absorption systems that might contaminate and prematurely truncate the proximity zone regions. 

In \citetalias{Eilers2020} we conducted a multi-wavelength survey of these quasars ranging from sub-mm, optical to infrared observations, in order to increase the precision on their proximity zone measurements. We determined the systemic redshift of these quasars based on their \cii\ $158\,\mu$m or $\rm CO(6-5)$ $3$~mm emission lines observed with the Atacama Large Millimetre Array (ALMA) or the NOrthern Extended Millimeter Array (NOEMA) at the Institute de Radioastronomie Millim{\'e}trique (IRAM), since uncertainties in the systemic redshift constitute the largest source of uncertainty in $R_{\mathrm p}$ measurements \citep{Eilers2017a}. These sub-mm and far-infrared (FIR) lines provide the most robust estimates of the systemic redshift since they arise from the cold gas reservoir within the quasar's host galaxy itself, whereas rest-frame UV emission lines might be displaced from the systemic redshift due to strong internal motions or winds in the broad line region of the quasars \citep[e.g.][]{Richards2002, Meyer2019, Schindler2020}. 
Nevertheless, we still need to account for small systematic uncertainties on the systemic redshifts even when estimated based on sub-mm or FIR emission lines. Thus, we conservatively estimate a systematic uncertainty of the redshift estimate of $\Delta v\approx 100\,\rm km\,s^{-1}$, i.e. $\Delta z\approx0.0023$ at $z\approx6$, which is an uncertainty comparable to the kinematics of the host galaxies \citep[e.g.][]{Neeleman2019, Venemans2019}. We ignore statistical uncertainties in the emission line peak ($\Delta v\lesssim 10\,\rm km\,s^{-1}$), as they are much smaller than the systematic uncertainty.
This redshift uncertainty corresponds to a systematic uncertainty on the proximity zone of $\Delta R_{\mathrm p}\approx0.14$~pMpc for a quasar at $z\approx6$. 

For quasars without available sub-mm observations or non-detections we estimated the systemic redshift from the \mgii\ broad emission line. However, this line is systematically blueshifted compared to the systemic redshift of the quasar \citep[e.g.][]{Mazzucchelli2017, Venemans2016, Meyer2019, Schindler2020} and thus we account for these systematic uncertainties and shift the redshift estimate by $\Delta v ($\mgii-\cii$) = -391^{+256}_{-455}\,\rm km\,s^{-1}$ to the systemic frame \citep{Schindler2020}. 

The optical and near-infrared (NIR) spectra in our data sample were observed with the X-Shooter echelle spectrograph on the Very Large Telescope (VLT) and the DEep Imaging Multi-Object Spectrograph (DEIMOS) on the Keck II telescope. The spectra were reduced, combined, and flux-calibrated using \texttt{PypeIt} \citep{Prochaska2020}, a newly developed algorithm for semi-automated reduction of astronomical spectroscopic data. 

In order to calculate the sizes of the proximity zones $R_{\mathrm p}$ we normalize the spectra by their continuum emission, which is estimated by means of a principal component analysis (PCA) trained on low-redshift quasar spectra \citep[e.g.][]{Suzuki2006, Paris2011, Davies2018, Bosman2020}. The coefficients of the principal components for the continuum model are estimated on the unabsorbed part of the quasar spectra, i.e. redwards of the \lya\ emission line, and then projected onto a set of principal components bluewards of the \lya\ emission line, where the $z\gtrsim 6$ quasar spectra are affected by IGM absorption \citepalias[see][for details]{Eilers2020}.
The continuum-normalized flux is then smoothed with a $20${\AA}-wide (observed frame) boxcar function, which corresponds to $\sim 1$~pMpc or a $\sim 700\,\rm km\,s^{-1}$ window at $z=6$. The location at which the smoothed flux drops below the $10\%$ transmission level defines the end of the quasars' proximity zones \citep{Fan2006, Carilli2010, Eilers2017a}. In \citetalias{Eilers2020} we carefully search for any dense absorption systems such as Damped \lya\ Systems (DLAs) or Lyman Limit Systems (LLSs) in the close vicinity of the quasars to exclude a premature truncation of the proximity zones. Such absorption systems are not modeled in our simulations (see \S~\ref{sec:RT}) and thus need to be excluded from our sample for the lifetime analysis. 
To this end we stacked the spectral regions that would show metal absorption lines associated with a hypothetical LLS at the edge of the proximity zone and compared this stack to the composite spectrum of $20$ low-redshift LLSs from \citet{Fumagalli2013}. The authors of this study conclude that the gas giving rise to the composite LLS spectrum is likely metal-poor, i.e. [Fe/H]$\lesssim -1.5$. We did not find any evidence of such metal-poor LLSs truncating the quasar proximity zone sizes, although the presence of very highly metal-poor or pristine systems cannot be excluded. 

All measurements of the systemic redshifts and the proximity zone sizes are shown in Table~\ref{tab:tq}.

\begin{deluxetable*}{lCcLCC}[!t]
\setlength{\tabcolsep}{18pt}
\renewcommand{\arraystretch}{1.2}
\tablecaption{Quasar sample. \label{tab:tq}}
\tablehead{\colhead{object} & \dcolhead{z_{\rm em} (\pm \sigma_{\rm sys})}  & \dcolhead{z_{\rm line}}& \dcolhead{M_{1450}} & \dcolhead{R_{\mathrm p}} & \dcolhead{\log_{10}t_{\rm Q}}\\
\colhead{} & \colhead{} & \colhead{} & \colhead{} & \colhead{[pMpc]} & \colhead{[yr]}}
\startdata
PSO\,J004+17 & 5.8166\pm0.0023 & [\ion{C}{2}] & -26.01 & 1.16\pm0.15 & 3.6^{+0.5\,(+2.5)}_{-0.4\,(-1.0)} \\
PSO\,J011+09 & 6.4695\pm0.0025 & [\ion{C}{2}] & -26.85 & 2.42\pm0.13 & 5.0^{+2.0\,(+3.7)}_{-0.8\,(-1.2)} \\
VDES\,J0323--4701 & 6.250^{+0.011}_{-0.006} & \ion{Mg}{2} & -26.02 & 2.26^{+0.62}_{-0.35} & 5.6^{+2.0\,(+3.1)}_{-1.2\,(-1.9)}  \\
VDES\,J0330--4025 & 6.249^{+0.011}_{-0.006} & \ion{Mg}{2} & -26.42 & 1.69^{+0.62}_{-0.35} & 4.1^{+1.8\,(+4.3)}_{-0.9\,(-2.0)} \\
PSO\,J158--14 & 6.0685\pm0.0024 & [\ion{C}{2}] & -27.41 & 1.95\pm0.14 & 3.8^{+0.4\,(+3.2)}_{-0.3\,(-0.6)} \\
SDSS\,J1143+3808 & 5.8366\pm0.0023 & CO(6--5) & -26.69 & 3.93\pm0.63 & >4.8\,(>4.2)\\ 
PSO\,J261+19 & 6.494^{+0.011}_{-0.006} & \ion{Mg}{2} & -25.69 & 3.36^{+0.59}_{-0.33} & >5.7\,(>4.7)  \\
CFHQS\,J2100--1715 & 6.0806\pm0.0024 & [\ion{C}{2}] & -25.55 & 0.37\pm0.14 & 2.3^{+0.7\,(+1.7)}_{-0.7\,(-1.2)}  \\
CFHQS\,J2229+1457 & 6.1517\pm0.0024 & [\ion{C}{2}] & -24.78 & 0.47\pm0.14 & 2.9^{+0.8\,(+2.8)}_{-0.9\,(-1.6)}\\
PSO\,J359--06 & 6.1722\pm0.0024 & [\ion{C}{2}] & -26.79 & 2.83\pm0.14 & 5.2^{+1.9\,(+3.5)}_{-0.9\,(-1.3)} \\
 \enddata
\tablecomments{The columns show the name of the quasar, its systemic redshift estimate with its systematic uncertainty and the emission line it is derived from, its absolute magnitude $M_{1450}$, as well as the measurement of the size of the proximity zone. The last column shows the median and $16-84$th ($2.5-97.5$th) percentile lifetime estimates. Note that all lifetime estimates have a systematic uncertainty of $\sigma_{{\rm sys},\,\log_{10}t_{\rm Q}}\approx 0.4$ (see \S~\ref{sec:caveats} for details). }
\end{deluxetable*}

\section{Radiative Transfer Simulations}\label{sec:RT}

In order to interpret the proximity zone measurements we conduct a series of RT simulations of the effect of ionizing radiation emitted by the quasar, which ionizes the IGM along the line-of-sight \citep{BoltonHaehnelt2007a}. We apply the 1D RT code by \citet{Davies2016} to skewers from the cosmological hydrodynamical Nyx simulation within a box of $100\,{\rm Mpc}\,h^{-1}$ on a side \citep{Almgren2013, Lukic2015}. The simulation includes $4096^3$ baryonic (Eulerian) grid elements and dark matter particles and was designed for precision cosmology studies of the diffuse gas in the IGM. The RT code computes the time-dependent ionization and recombination rates of six particle species, i.e. $e^{-}$, \ion{H}{1}, \ion{H}{2}, \ion{He}{1}, \ion{He}{2}, and \ion{He}{3}, as well as photo-ionization heating from the quasar's radiation itself and the associated heating and cooling of the intergalactic gas due to the expansion of the universe and inverse Compton cooling off the cosmic microwave background (CMB). 
Intergalactic gas that is self-shielding from the UVB is treated via the method presented in \citet{Rahmati2013}, whereas the self-shielding to the quasar light is treated self-consistently in the RT simulations. 

The quasar's escape fraction of ionizing UV photons is assumed to be unity. This is consistent with measurements at lower redshift, where such measurements of the escape fractions are easily possible, since there is no or only very little Lyman series forest absorption. \citet{Stevans2014} for instance conducted a measurement for quasars at $z<1.5$ and found no break in the quasar spectra at the Lyman Limit with an optical depth $\tau<0.01$, i.e. $f_{\rm esc}>99\%$. At higher redshifts such measurements are more challenging due to the higher IGM absorption. However, \citet{Worseck2014} constructed stacked quasar spectra at $z\sim 5$ in order to measure the mean free path of Lyman continuum photons, which also do not show any break around the Lyman Limit, similarly indicating high escape fractions. 

We assume the luminosity--dependent bolometric correction from $M_{1450}$ given in Table~3 of \citet{Runnoe2012}, and convert $M_{1450}$ to ionizing luminosity using the spectral energy distribution (SED) by \citet{Lusso2015}. This SED assumes a spectral index of $\alpha_{\nu}=-1.7$. We discuss the influence of the chosen SED on our results in \S~\ref{sec:caveats}. 

For each quasar we take $1000$ skewers of density, temperature and peculiar velocity, which are drawn from the centers of the most massive dark matter halos in the Nyx simulation. In order to avoid interpolation errors we extrapolate $6$ sightlines from each halo along the $\pm x,y,z$-axes. The halos have masses between $4\times 10^{11}\,M_{\odot}\lesssim M_{\rm halo}\lesssim 3\times10^{12}\,M_{\odot}$, however, our results are nearly independent of the chosen host dark matter halo masses \citep[see also][]{Keating2015}. We have Nyx simulation outputs at $z=6$ and $z=6.5$ available, from which we choose the output closest to the quasar's redshift $z_{\rm em}$ and re-scale the physical densities by $(1+z_{\rm em})^3$. 

The ionization state of the IGM before the quasar turns on is given by post-reionization conditions and thus considered to be highly ionized, i.e. with a neutral fraction of $x_{\rm H\,I}\sim 10^{-4}$. Furthermore, we assume an ultraviolet background (UVB) radiation with an ionization rate of $\Gamma_{\rm UVB}=2\times 10^{-13}\,\rm s^{-1}$, which is consistent with observations of the \lya\ forest opacity \citep[][]{Fan2006, WyitheBolton2011, Davies2018a, Yang2020b}. 
Note that our Nyx simulation does not model the inhomogeneous reionization process, but rather assumes that reionization happens instantaneously at $z\sim 10$, i.e. reionization was completed long before the redshifts of the quasars considered in this paper. 
We discuss the influence of this assumption on our results in more detail in \S~\ref{sec:caveats}. 

Furthermore, Nyx does not resolve dense self-shielding gas structures, that would result in DLAs or LLSs and could prematurely truncate the proximity zones. For this reason we exclude any quasar spectra that exhibit such absorption features from our analysis \citepalias{Eilers2020}. Recently, \citet{ChenGnedin2020} have analyzed the extent of proximity zones using the Cosmic Reionization On Computers (CROC) simulations, which have sufficiently high spatial resolution to resolve such structures. Their study confirms that overdensities of $\Delta\gtrsim10^3$ within $6$~pMpc distance from the quasar corresponding to DLAs and LLSs can cause small proximity zone sizes despite long quasar lifetimes and hence quasar sightlines exhibiting these features would result in biased lifetime estimates.

\subsection{Evolution of the Proximity Zone Sizes with Quasar Lifetime}\label{sec:Rp_evol}

In order to study the evolution of proximity zone sizes with the quasar's lifetime, we assume a simple light--bulb model for the quasars' light curves, i.e. the quasars turn on abruptly and emit at a constant luminosity during its entire lifetime. We compute RT outputs in steps of $\Delta \log_{10}(t_{\rm Q}/\rm yr) =0.1$ from $t_{\rm Q,\,min}=10$~yr up to the age of the universe, i.e. $t_{\rm Q,\,max}\sim10^{8.9}$~yr at the redshift of the quasars. 

\begin{figure*}[!t]
\centering
\includegraphics[width=.92\textwidth]{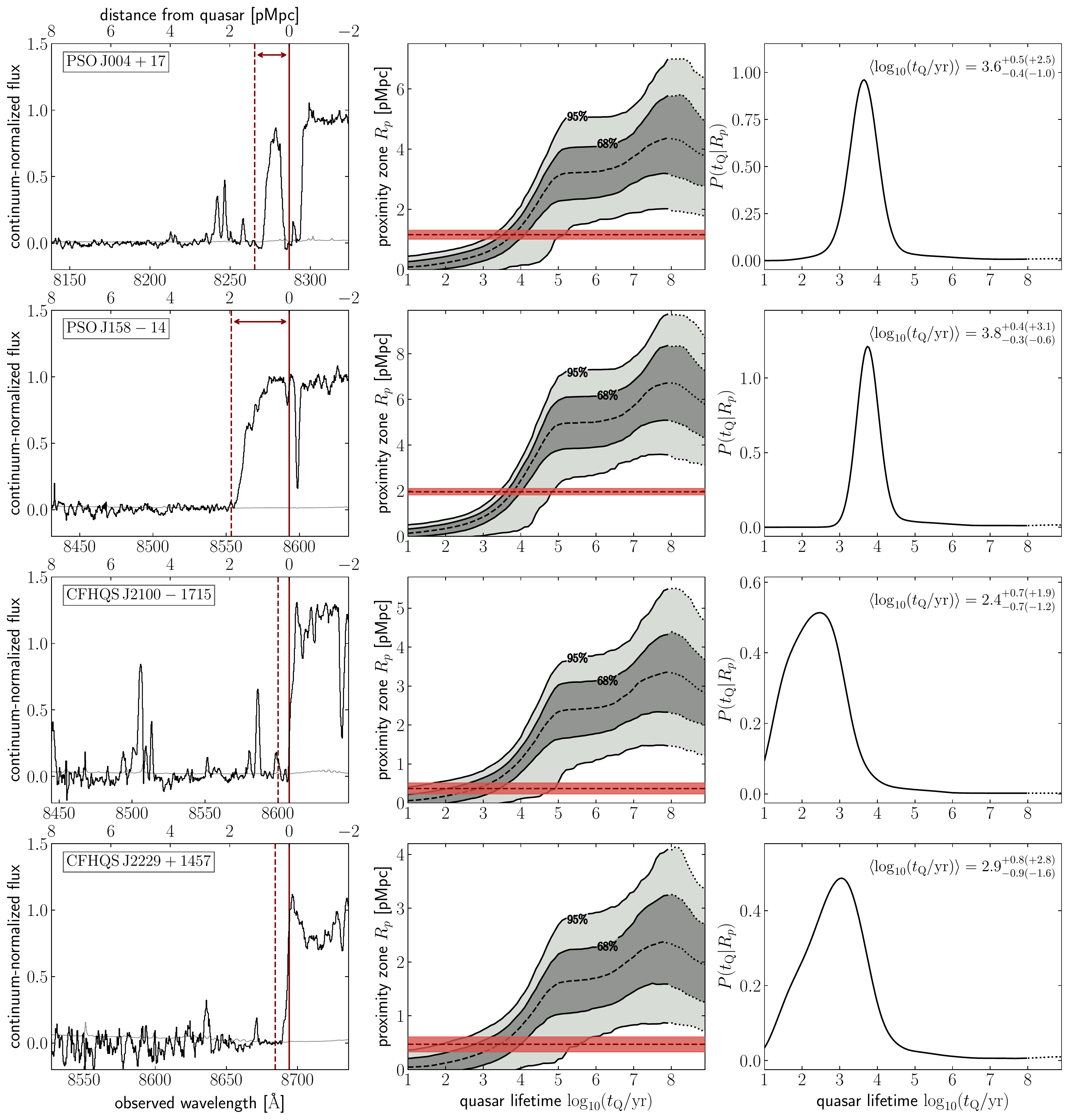}
\caption{Lifetime estimates for the four young quasars in our sample. \textit{Left panels:} Continuum-normalized quasar spectra around the \lya\ emission line (black) and their noise vector (grey), which have been inverse--variance smoothed with a $3$ pixel filter. The proximity zones extend between the systemic redshift estimate of the quasars shown by the red solid lines, and the red dashed lines indicating the end of the proximity zone. \textit{Middle panels:} Evolution of proximity zone sizes with quasar lifetime from RT simulations. The proximity zone measurements and their $1\sigma$ ($68$th percentile) uncertainties are indicated by the red dashed lines and shaded regions. \textit{Right panels:} Posterior probability distribution of the quasar's lifetime marginalized over the proximity zone measurement uncertainty. The dotted lines in the middle and right panels indicate that the proximity zone size evolution beyond $\log_{10}(t_{\rm Q}/\rm yr) > 8$ should be taken with caution due to the limitations in our modeling procedure described in \S~\ref{sec:Rp_evol}. \label{fig:young_qsos}} 
\end{figure*}

\begin{figure*}[!t]
\centering
\includegraphics[width=.92\textwidth]{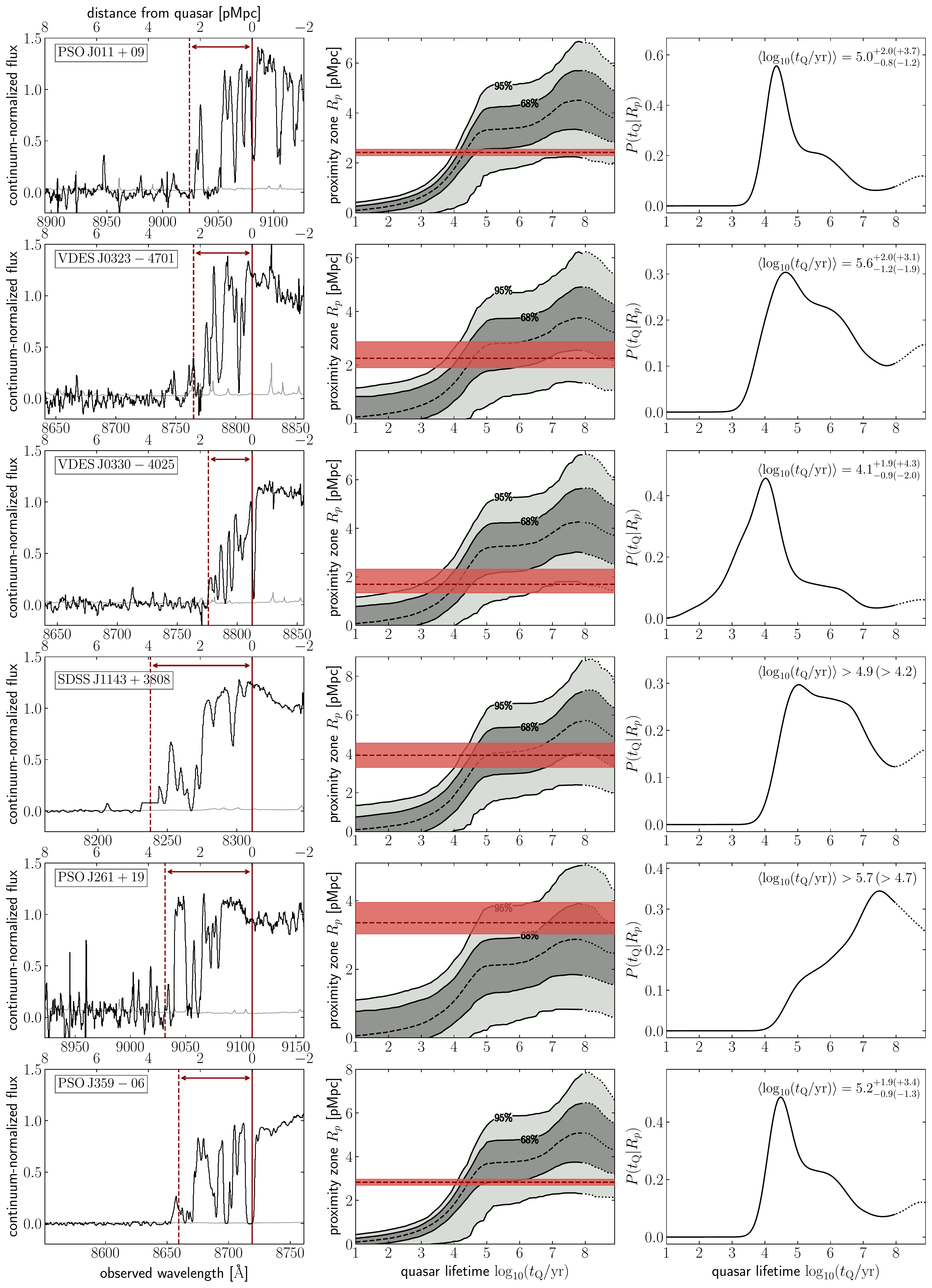}
\caption{Same as Fig.~\ref{fig:young_qsos} for the remaining quasars in our sample.\label{fig:remaining_qsos}} 
\end{figure*}

To measure the size of the proximity zones, we apply the same method to the simulated quasar sightlines as to the continuum--normalized quasar spectra as described in \S~\ref{sec:sample}. The dependence of simulated proximity zones on the quasar's lifetime is shown in the middle panels of Fig.~\ref{fig:young_qsos} and \ref{fig:remaining_qsos}. The overall dependence is the same for each quasar, however, the size of $R_{\mathrm p}$ at a given lifetime depends on the quasar's absolute magnitude, i.e. the ionizing luminosity of the quasar. 

For short quasar lifetimes, the size of the proximity zones increases when the quasar's radiation ionizes the surrounding IGM due to the finite response time of the intergalactic gas to the quasar's radiation (see middle panels of Fig.~\ref{fig:young_qsos} and \ref{fig:remaining_qsos}). Once the gas reaches ionization equilibrium with the quasar's radiation at around $t_{\rm eq}\approx \Gamma_{\rm H\,I}^{-1}(R_{\mathrm p})\approx 3\times 10^4$~yr, where $\Gamma_{\rm H\,I}(R_{\mathrm p})$ denotes the combined photoionization rate from the quasar and the UVB at $R_{\mathrm p}$, the proximity zone ceases to grow further \citep{Khrykin2016, Eilers2017a, Davies2019a}. 
For longer lifetimes of $t_{\rm Q}\gtrsim 10^7$~yr the proximity zone increases again as a result of the thermal proximity effect, which is due to the heating of the IGM around quasars when their radiation doubly-ionizes helium \citep{Meiksin2010, Bolton2012, Khrykin2017}. 
Once the quasar lifetime exceeds $t_{\rm Q}\gtrsim 10^8$~yr we observe a mild decrease in $R_{\mathrm p}$ again, which is caused by the cooling of the IGM due to Compton cooling off the CMB and the expansion of the universe. The intergalactic gas around the quasar is now completely ionized and cannot be heated further from the quasar's radiation. Therefore the temperature $T$ of the IGM decreases, and hence the recombination rate $\alpha_{\rm HII}$ increases, i.e. $\alpha_{\rm HII}\propto T^{-0.7}$, which mildly reduces the size of the proximity zones. 

However, on these long timescales exceeding $>10^8$\,yr, the accuracy of our fixed-time assumption (i.e. that the redshift of the RT is fixed) degrades in two ways: First, collisional rates in the simulation (e.g. collisional cooling, recombination) will have been on average underestimated at times prior to a $>10^8$\,yr snapshot, due to the evolving mean density of the IGM.
Second, the long-term equilibrium thermal state of the gas in the simulations becomes slightly incorrect because we neglect adiabatic heating and cooling from structure formation \citep[see e.g.][]{McQuinn2015}. Both of these effects should have only a minor influence on the evolution of $R_{\mathrm p}$. However, due to these shortcomings of the RT simulations we will consider lifetime estimates exceeding $>10^8$\,yr with caution.

\section{Quasar Lifetime Estimates}\label{sec:lifetimes}

In order to estimate the lifetime of the quasars in our data sample we compare their proximity zone measurements to the simulated proximity zones at different lifetimes. To this end, we first forward-model the uncertainty on the proximity zone measurements onto the simulated sightlines by adding random draws from a normal distribution with $\sigma = \Delta R_{\mathrm p}$ to the simulated proximity zone sizes. This broadens the distribution of $R_{\mathrm p}$ at any given lifetime according to the measurement uncertainty. 
Note that quasars whose redshift estimates are based on the broad \mgii\ emission line and thus were shifted to the systemic redshift frame, have an asymmetric uncertainty on their proximity zone size estimates. For those objects we randomly draw from the measured velocity offsets between the \mgii\ and \cii\ emission lines of $28$ $z\gtrsim6$ quasars in \citet{Schindler2020}, which changes each simulated proximity zone measurement by $\Delta R_{\mathrm p} = \Delta v/H(z)$, where $H(z)$ denotes the Hubble parameter at the quasars' redshifts. 

The likelihood is $\mathscr{L} = P(R_{\mathrm p}|t_{\mathrm Q})$ and we take a flat logarithmic prior on $t_{\mathrm Q}$. 
We perform a 2D kernel density estimation (KDE) on the distribution of $R_{\mathrm p}$ as a function of $\log_{10}(t_{\rm Q})$ \citep[see also][]{Khrykin2019}, and evaluate the 2D distribution at $R_{\mathrm p}$ to obtain a posterior probability distribution function for each quasar's lifetime marginalized over the proximity zone size, i.e. $P(t_{\rm Q}|R_{\mathrm p})$. 
The posterior probability distribution functions for the lifetime of each quasar are shown in the right panels of Fig.~\ref{fig:young_qsos} and \ref{fig:remaining_qsos}. We take the median of this posterior probability distribution as the best lifetime estimate with uncertainties given by the $68$th ($95$th) percentile.

Note that we adopt a lower (or upper) limit on the lifetime estimate if the probability exceeds an arbitrarily chosen threshold of $P>0.15$ at the boundaries of the considered range of lifetime values, which are set by our chosen prior. 
The lower boundary of our prior, i.e. $t_{\rm Q} \geq 10$~yr, is set because most of the analyzed quasars are known for $\gtrsim10$~yr, whereas the upper boundary of our prior, i.e. $t_{\rm Q} \leq 10^{8.9}$~yr, approximately corresponds to the age of the universe at the quasars' redshifts. 
Table~\ref{tab:tq} shows the lifetime estimates for all quasars in our sample.

We find four quasars in our data sample for which we estimate a very short lifetime, i.e. $t_{\rm Q}< 10^4$~yr (Fig.~\ref{fig:young_qsos}). For the remaining six quasars our estimates indicate longer lifetimes of $t_{\rm Q}\gtrsim 10^5$~yr (Fig.~\ref{fig:remaining_qsos}). These quasars ended up in our data sample despite the pre-selection for potentially young quasars because their preliminary redshift estimate based on template fitting were scattered towards lower redshifts and thus smaller proximity zones. The more precise estimates based on sub-mm emission lines or the \mgii\ emission line that we obtained in \citetalias{Eilers2020} reveal a higher systemic redshift, and thus their proximity zones are larger and their lifetime estimates longer. 


\subsection{Sources of Systematic Uncertainty}\label{sec:caveats}

Various sources of uncertainty in our modeling procedure can affect our results, which we will discuss now. For instance, the choice of SED affects the number of emitted ionizing photons from the quasar. Our fiducial SED assumes a spectral index of $\alpha_{\nu}=-1.7$ \citep{Lusso2015}. In order to estimate the influence of the SED on our results, we alter the spectral index by $\pm \Delta\alpha_{\nu}=0.5$, i.e.\ $\alpha_{\nu}=-1.2$ and $\alpha_{\nu}=-2.2$, and estimate the quasars' lifetimes for the modified SED. This alters our lifetime estimates slightly and thus the systematic uncertainty on the lifetime estimates introduced by the choice of the quasars' SED is $\Delta \log_{10}t_{\rm Q}\lesssim 0.3$ (see Fig.~\ref{fig:SED} in Appendix~\ref{abs1}). 

Furthermore, the choice of the ionizing background $\Gamma_{\rm UVB}=2\times10^{-13}\rm s^{-1}$ also influences the quasar lifetime estimates. \citet{Davies2018b} have shown that patches with $\sim 4$ times lower $\Gamma_{\rm UVB}$ can exist at $z\sim 6$. To this end we evaluate the influence that a weaker ionizing background of $\Gamma_{\rm UVB}=4\times10^{-14}\rm s^{-1}$ would have on our results and find that it introduces a systematic uncertainty of $\Delta \log_{10}t_{\rm Q}\approx 0.2$. Note, however, that the regions around quasars within a few pMpc are likely to have a relatively high UVB compared to the mean UVB \citep{Davies2020_ghost}, in which case our choice of the mean UVB would be conservative. 
Thus the combined, i.e.\ added in quadrature, systematic uncertainties on the quasar lifetime estimates arising from the choice of SED as well as the ionizing background are $\sigma_{{\rm sys},\,\log_{10}t_{\rm Q}}\approx0.4$. 

Some recently published late reionization models predict that $\gtrsim 30\%$ of the cosmic volume is occupied with neutral gas fractions of $\langle x_{\rm HI}\rangle \gtrsim 0.1$ as late as $z\sim6$ \citep{Kulkarni2019, Keating2020}. If the quasars at $z\sim 6$ would indeed still reside in large neutral patches, this could potentially influence our analysis. However, if the intergalactic gas surrounding the quasars is indeed still mostly neutral, it should imprint a damping wing on the \lya\ emission line \citep{MiraldaEscude1998}. 
While such damping wings have to date only been detected in quasars at $z\gtrsim 7$ \citep{Greig2017, Davies2018, Wang2020, Yang2020a}, we cannot rule out securely the existence of highly neutral gas around the young quasars. 
Nevertheless, a still partially incomplete reionization process with remaining neutral patches, fluctuations in the IGM temperature or a varying mean free path of ionizing photons may affect the proximity zones, but such modeling will be part of future work.

\section{Comparison to Previous Studies}\label{sec:comparison}

Many other studies have set constraints on the timescales of quasar activity with a variety of different methods. We will summarize some of the results from the literature and distinguish between constraints obtained by analyzing the proximity zones in the observed spectra along the line-of-sight (\S~\ref{sec:line-of-sight}), by studying the ionization echo of the quasars' radiation (\S~\ref{sec:echo}), or constraints on the duty cycle of quasars from clustering studies (\S~\ref{sec:clustering}). 

\subsection{Lifetime Constraints from Quasar Emission along the Line-of-Sight}\label{sec:line-of-sight}

A similar approach to obtain quasar lifetime estimates to the one presented in this work has been applied to quasars at $z\gtrsim 7$  \citep{Davies2019b}. While the IGM at $z\sim 6$ is already highly ionized, at $z\gtrsim 7$ it still contains a significant neutral gas fraction since the Epoch of Reionization is not yet completed. Thus, quasar spectra at $z\gtrsim 7$ exhibit a damping wing around the \lya\ emission line due to the increased neutral gas fraction of the IGM \citep{MiraldaEscude1998}. The shape and strength of these damping wings provide information about the total number of ionizing photons that have been emitted into the IGM, which can be used to obtain estimates on the integrated lifetime, i.e. the quasars' duty cycle. Applying this method to four quasars at $z\gtrsim 7$ resulted in estimates of $t_{\rm Q}\lesssim1$~Myr \citep[][Davies et al. in prep.]{Davies2019b, Wang2020, Yang2020a}. 

Quasars at $z\sim 3-4$ exhibit a proximity zone in the \heii\ \lya\ forest \citep[e.g.][]{Hogan1997}, which provides constraints on the lifetime of quasars similarly to the proximity zones in the \ion{H}{1} \lya\ forest at $z\sim 6$. However, due to the lower photoionization rate at the \ion{He}{2} ionizing edge \citep{Shull2004}, the timescale to obtain ionization equilibrium is longer, i.e. $t_{\rm eq,\,He\,II}\sim 3 \times 10^7$~yr, and thus the extent of the proximity zone remains sensitive to longer quasar lifetimes before reaching ionization equilibrium \citep{Khrykin2016}. \citet{Khrykin2019} applied this method to obtain lifetime constraints for six quasars at $z\sim4$, for which they estimate lifetimes of $t_{\rm Q}\sim 10^6-10^7$~yr \citep[see also][]{Worseck2021, Khrykin2021}. 

An estimate of the \textit{average} effective lifetime of the $z\sim 6$ quasar population has been obtained by stacking the proximity zones of a luminosity-selected, i.e. unbiased towards young ages, sample of $15$ quasars (Morey et al. subm.). By comparing the stacked transmitted flux profile around quasars, selected solely based on their absolute magnitude, i.e. $-26.6 > M_{1450} > -27.4$, and systemic redshifts, i.e. $5.8 < z < 6.5$ based on sub-mm emission lines, to the stacked profile of simulated data sets from RT simulations at different lifetimes, they obtain a lifetime estimate for the whole quasar population of $\log_{10}(t_{\rm Q}/{\rm yr})=5.7^{+0.5}_{-0.3}$. 

Finally, \citet{Davies2020} performed a similar analysis to ours on the proximity zone of the hyperluminous $z\approx6.3$ quasar SDSS\,J0100+2802, one of the young quasars found in \citet{Eilers2017a}, to test the hypothesis that it could be strongly gravitationally lensed \citep{Fujimoto2020}. 
Based on the size and shape of the quasar's proximity zone they exclude a strong magnification and confirm the young age of the object $\log_{10}(t_{\rm Q}/{\rm yr})=4.28^{+0.61}_{-0.15}$. 

\begin{figure*}[!t]
\centering
\includegraphics[width=\textwidth]{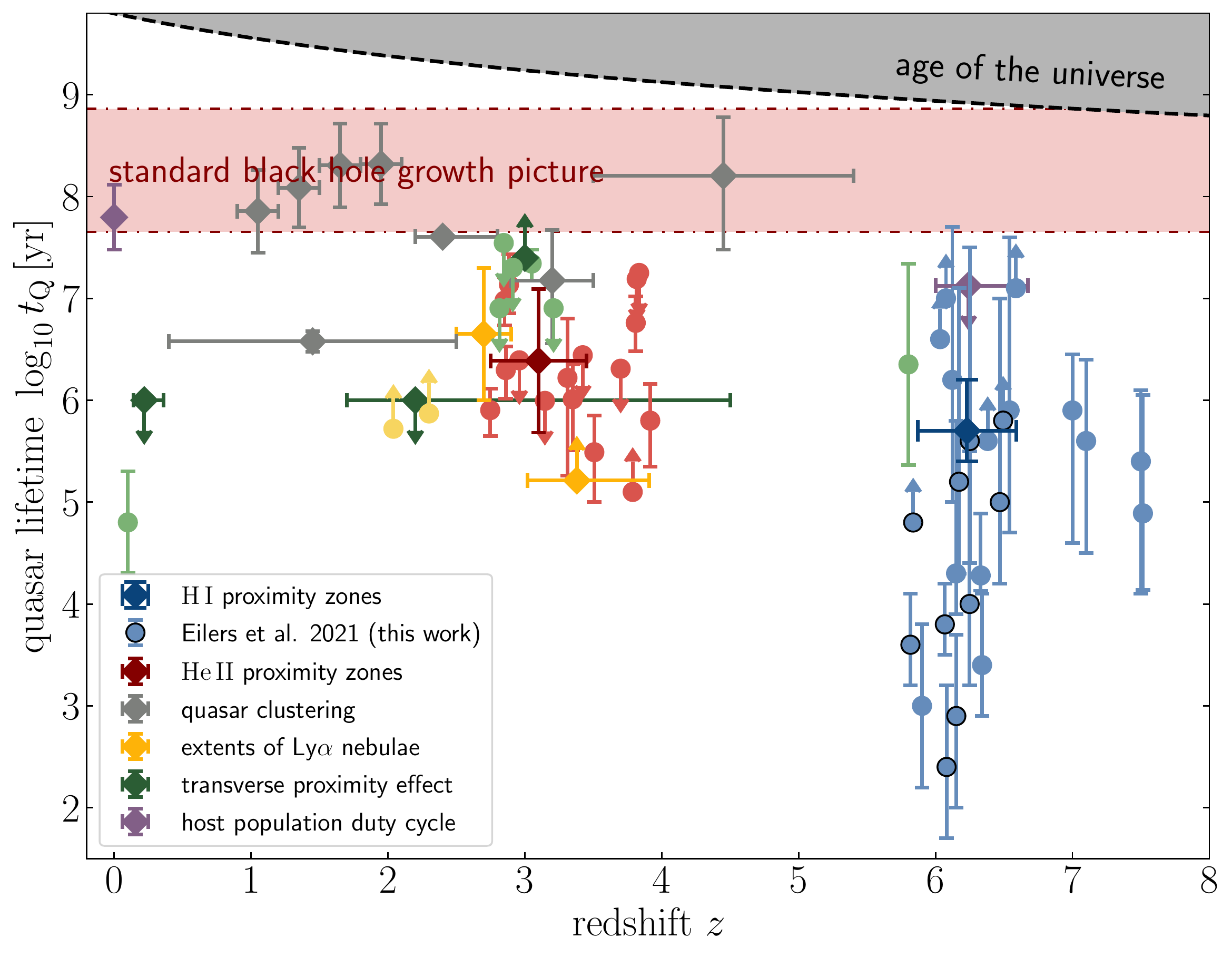}
\caption{Comparison of quasar lifetime estimates from the literature as a function of redshift. 
Light colored points show measurements from individual objects, whereas the darker colored diamonds show quasar population averages. Studies based on the quasars' \ion{H}{1} proximity zones along the line-of-sight are shown in blue (individual measurements taken from this work are shown as data points with black outline, plus additional measurements from \citet{Eilers2018b}, \citet[][in prep.]{Davies2019b, Davies2020} and  \citet{Andika2020}, population average from Morey et al. subm.), whereas measurements based on the \ion{He}{2} proximity zones are shown in red (individual measurements taken from \citet{Khrykin2019} and \citet{Worseck2021}, population average from \citet{Khrykin2021}). Constraints on the quasars' duty cycle from clustering studies are shown in grey \citep{Shen2007, Shankar2010b, White2012, Laurent2017}, and the extents of \lya\ nebulae provide quasar lifetime constraints depicted in yellow (individual measurements from \citet{Cantalupo2014} and \citet{Hennawi2015}, averages from \citet{TrainorSteidel2013} and \citet{Borisova2016}). Furthermore, quasar lifetimes that have been inferred from the transverse proximity effect are shown in green (individual measurements from \citet{Keel2012}, \citet{Schmidt2018}, and \citet{Bosman2019}, population average from \citet{KirkmanTytler2008}, \citet{Schmidt2017}, and \citet{Oppenheimer2018}), and the duty cycle of the quasar host population shown in purple has been inferred by \citet{YuTremaine2002} and \citet{Chen2018}. 
The red shaded area shows the range of lifetimes expected from the standard black hole growth picture, i.e. $\sim 1-16$ Salpeter times, assuming a fiducial radiative efficiency of $\epsilon=0.1$ and Eddington-limited accretion rates (see Eqn.~\ref{eq:salpeter}).  \label{fig:comp}}
\end{figure*}

All lifetime constraints based on \ion{H}{1} or \ion{He}{2} proximity zones measured along the line-of-sight are shown in Fig.~\ref{fig:comp} as blue and red shaded data points, respectively. 

\subsection{Lifetime Constraints based on the Quasars' Ionization Echo}\label{sec:echo}

The time lag between the onset of a quasar's ionizing radiation and its arrival at a nearby galaxy or its crossing of a background sightline can also be used to constrain the duration of the quasar's emission, as well as its geometry and nuclear obscuration. 

For instance, \citet{Schmidt2018} studied the \heii\ transverse proximity effect, i.e. the enhanced flux transmission in the \heii\ \lya\ forest in a spectrum of a background quasar due to the ionizing radiation of a quasar in the foreground \citep[see also][]{Jakobsen2003}. From the analyzed objects however, only one shows a clear transmission spike allowing constraints on the quasar age, but for three other ones they do not detect transmission, from which they conclude that these quasars could either be young, i.e. $t_{\rm Q}\lesssim 10$~Myr or highly obscured \citep[see also][]{Schmidt2017}. 
Similarly, \citet{KirkmanTytler2008} study the transverse proximity effect in the \ion{H}{1} \lya\ forest by means of close quasar pairs at $z\sim 2.2$ and conclude that quasar lifetimes of less than a million years can explain the absorption pattern they observe in the spectra of the background quasars at the location of the foreground objects. 

\citet{Bosman2019} recently set constraints on the lifetime of a quasar at $z\approx5.8$ by studying the emission line profiles from nearby \lya--emitting galaxies (LAEs). The observed line profiles are expected to be double--peaked if the galaxies are embedded in a highly ionized IGM, whereas if the quasar's radiation has not yet reached their location and the IGM is less ionized, the blue peak would be absorbed. By modeling the time lag between the quasar and the surrounding galaxies they constrain the lifetime of the quasar to be $2.3\times10^5\,{\rm yr}\lesssim t_{\rm Q}\lesssim 2.2\times10^7$~yr. 

A similar study by \citet{TrainorSteidel2013} investigated the fluorescent emission of LAEs due to reprocessing of the ionizing radiation from a nearby quasar. Their study of eight hyperluminous quasars at $2.5<z<2.9$ allows joint constraints on the geometry and history of the emitted ionizing radiation, from which they estimate an average lifetime of $10^6\,{\rm yr}\lesssim t_{\rm Q}\lesssim 2\times 10^7$~yr. 

Lower limits on the lifetime of quasars can be derived from the extended \lya nebulae, since continuous emission is required to sustain the nebular emission found ubiquitously around $z\sim 2-4$ quasars \citep{Hennawi2013, Borsiova2016_nebulae, FAB2019} and around a majority of $s\sim6$ quasars \citep{Farina2019, Drake2019}. Giant nebulae extending up to $450$~kpc distance from the quasar hosts provide evidence for sustained quasar lifetimes $\gtrsim 1$~Myr \citep{Cantalupo2014, Hennawi2015}. 

At very low redshifts around $z\lesssim0.1$ the discovery of extended highly ionized nebular regions around quiescent galaxies, such as ``Hanny’s Voorwerp'' \citep{Lintott2009}, has lead to constraints on the timescales of quasar activity around currently in--active, quiescent objects. The presence of high-ionization emission lines suggests photoionized gas from an active galactic nucleus (AGN), but the optical spectrum does not reveal a currently active nuclear core. Thus, \citet{Keel2012} argue that the AGN luminosity must have dropped significantly in the recent past, and that the observed ionized nebula manifests as the light echo of the faded AGN. Based on the extents of the ionized regions and the number of faded to non-faded AGN they estimate that the nuclear active phase in these galaxies must have lasted for $0.2-2\times 10^5$~yr \citep[see also][]{Schawinski2015, Sartori2018}. 

A related argument has been brought forward by \citet{Oppenheimer2018}, who simulate the highly ionized circumgalactic medium around $z\sim0.2$ galaxies, to reproduce the strong abundance of \ion{O}{6} absorbers observed in the COS-Halos survey \citep{Tumlinson2011}. They argue that these galaxies likely hosted an AGN in their past with a lifetime of $t_{\rm Q}\lesssim 10^6$~yr, which is shorter than the recombination time of \ion{O}{6}, i.e. $t_{\rm rec}\sim 10^7$~yr, to explain the observed high \ion{O}{6} abundance. However, other studies have explained the abundance of such systems without invoking AGN photoionization \citep[e.g.][]{McQuinnWerk2018, Stern2018}. 

Constraints on the lifetime of quasars based on the light echo of their ionizing radiation are shown in Fig.~\ref{fig:comp} as green and yellow shaded data points.

\subsection{Constraints on the Quasar Duty Cycle from Clustering}\label{sec:clustering}

The duty cycle of quasars can be estimated from the ratio of the number density of dark matter halos hosting an active black hole to the total number of halos that could host quasars within the luminosity range of a given sample. Since more massive halos have a higher clustering bias \citep{Kaiser1984, Tinker2010}, the observed clustering of quasars determines the characteristic mass of their host halos. The abundance of such halos compared to the number density of quasars they host results in an estimate of the fraction of time that galaxies shine as luminous quasars \citep[e.g.][]{EfstathiouRees1988, HaimanHui2001, MartiniWeinberg2001, White2008, ConroyWhite2013}.

This approach to infer the duty cycle of quasars by measuring their number density and clustering strength has been adopted by several groups making use of the large quasar samples at $2\lesssim z\lesssim 4$, such as the Sloan Digital Sky Survey \citep[SDSS;][]{Shen2007, Shen2009, Shankar2010b},
the Baryon Oscillation Spectroscopic Survey \citep[BOSS;][]{White2012}, and most recently the extended--BOSS (eBOSS) sample \citep{Eftekharzadeh2015, Laurent2017}. These results are broadly consistent with duty cycles of $\sim 10^7-10^8$~yr, apart from \citet{Shankar2010b} who infer a shorter duty cycle for lower redshift quasars at $0.4\leq z\leq2.5$, although their measurements do not securely exclude longer duty cycles. 

\citet{YuTremaine2002} use the velocity dispersion of early--type galaxies in SDSS to estimate the local black hole mass density, which they find to agree with the black hole mass density accreted during optically luminous quasar phases estimated from quasar luminosity functions. 
They obtain an estimate of the duty cycle of quasars of $\sim (3-13) \times 10^7$~yr, which is comparable to the Salpeter time (Eqn.~\ref{eq:salpeter}).

A slightly different approach to measure the duty cycle of quasars at $z\sim 6$ was recently attempted by \citet{Chen2018} who made use of the \cii\ gas dynamics in quasar host galaxies observed with ALMA \citep{Decarli2018}, from which they estimate the dark matter halo masses of quasar hosts. 
Unfortunately, their results are strongly dependent on the underlying assumptions of how the \cii\ emitting gas populates the dark matter halos, but they derive an upper limit on the quasar duty cycle of $\lesssim 10^{7.1}$~yr at $z\sim 6$. 

In Fig.~\ref{fig:comp} the constraints from studies of the duty cycle of quasars and their host galaxies are shown as grey and purple data points.

\section{Discussion}\label{sec:discussion}

\subsection{Implications of Short Quasar Lifetimes}

The comparison to previous work reveals that multiple studies point towards an \textit{average} lifetime of $t_{\rm Q}\sim 10^6$~yr for the quasar population at large. 
While many of the aforementioned studies constrain the duty cycle and lifetime of an \textit{ensemble} of quasars, inferring quasar lifetimes from the extents of proximity zones offers the possibility to infer the lifetime of \textit{individual} objects. 

In this paper we confirmed the short lifetimes of four quasars, which increases the number of confirmed young quasars to seven in total, keeping in mind the three previously identified young objects \citep{Eilers2017a, Eilers2018b, Davies2020, Andika2020}. This confirms the fraction of young quasars within the whole quasar population to be $5\%\lesssim f_{\rm young}\lesssim 10\%$ reported in \citetalias{Eilers2020}. 
Thus when assuming an average lifetime of the quasar population is $t_{\rm Q}\sim 10^6$~yr, we expect to see quasars with short lifetimes of $t_{\rm Q}\sim 10^4-10^5$~yr $\sim 1-10\%$ of the time when randomly sampling a light--bulb light curve with $t_{\rm Q}\sim 10^6$~yr. Given the statistical error on the small sample size, this implies that the estimated fraction of young quasars $f_{\rm young}$ is approximately consistent with an average lifetime of the overall quasar population of $t_{\rm Q}\sim 10^6$~yr, which various studies agree on to be the preferred value at different redshifts ranging from $3.0\lesssim z \lesssim 7.5$ \citep[e.g.][Morey et al. subm.]{Khrykin2016, Davies2019b, Khrykin2021}. 

It is interesting to note that the fraction of confirmed young quasars within the quasar population as a whole is $\sim4\%$ for objects at $z<6.1$ and $\sim6\%$ at $z>6.1$. Keeping in mind the still preliminary statistics due to the small total number of confirmed young quasars, this suggests that there is no significant redshift evolution of the young quasar fraction within the probed redshift range. 

Assuming light--bulb light curves, the inferred short average quasar lifetime of $t_{\rm Q}\sim 10^6$~yr poses significant challenges to current models for the formation and growth of SMBHs in the center of the quasars' host galaxies. The quasars' optical and NIR spectra reveal that most observed luminous $z\sim6$ quasars host SMBHs of $\sim 10^9\,M_{\odot}$ \citep[e.g.][]{Mazzucchelli2017, Schindler2020}. Based on arguments laid out in \S~\ref{sec:intro}, these quasars must have been accreting for $t_{\rm Q} > 700$~Myr, when assuming a radiative efficiency of $\epsilon\sim0.1$ (see red shaded region in Fig.~\ref{fig:comp}). 
Thus, in order to reduce the Salpeter time to comply with the shorter observed average quasar lifetimes of $t_{\rm Q}\sim 10^6$~yr, our results might provide evidence for highly radiatively inefficient accretion rates, i.e. $\epsilon\sim 0.001$, as expected for ``super-Eddington'' or ``hyper-Eddington'' accretion disks \citep[e.g.][]{Inayoshi2016, BegelmanVolonteri2017}. 
The tension between the expected and observed quasar lifetimes can further be alleviated when assuming flickering quasar light curves with multiple episodes of quasar accretion instead of light-bulb light curves, as discussed in \S~\ref{sec:flickering}. 

An alternative explanation for the existence of SMBHs in high--redshift quasars despite their short UV luminous lifetimes could be that a majority of the black hole growth happens in highly obscured, dust--enshrouded environments \citep[e.g.][]{YuTremaine2002, Hopkins2005a, Hopkins2008, Eilers2018b, Davies2019b}. This latter scenario would imply the presence of a large fraction of obscured quasars in the high--redshift universe that has not yet been observed \citep[but see][]{Vito2018, Vito2019}.

\subsection{Effects of Flickering Quasar Light Curves}\label{sec:flickering}

The proximity zones of quasars are mostly sensitive to the most recent UV luminous episode of quasars, and thus some of the tension between the short lifetimes of the quasars and the time required to grow the SMBHs can be alleviated by invoking flickering quasar light curves, which would allow for multiple episodes of quasar activity and black hole growth. In this picture the UV emission from the quasar fluctuates either due to intrinsic variations in the accretion flow or time-variable obscuration along the line-of-sight. 

If a quasar's ionizing radiation fades, the intergalactic gas recombines and the neutral gas fraction increases. 
While the recombination timescale of hydrogen from a highly ionized to a completely neutral state is long, i.e. comparable to the age of the universe, the timescale on which the IGM can recombine from a \lya--transparent highly ionized state of $x_{\rm H\,I}\sim 10^{-5}$ to a \lya--opaque less ionized state of $x_{\rm H\,I}\sim 10^{-4}$ is much shorter. 
Although the size of the proximity zone depends non-trivially on the neutral gas fraction along the line-of-sight at the limiting optical depth of $\tau_{\lim}=2.3$, corresponding to a $10\%$ flux transmission in the \lya\ forest which defines the extent of $R_{\rm p}$, the time that it takes for the proximity zone of a $z\sim 6$ quasar to disappear after the quasar's ionizing emission has faded is similarly short, i.e. $\sim 10^4$~yr \citep[see][for details]{Davies2019a}. 

Furthermore, our model is limited to statements about the current luminosity of the quasar, which may not be representative of its average luminosity over the last equilibration timescale. If the quasars were to exhibit extreme variability with magnitude changes of $\sim 1$~mag within $\sim 15$ years, as observed in some quasars at low redshifts and lower luminosity \citep[e.g.][]{Rumbaugh2018}, our model would not capture these rapid changes appropriately. Moreover, the change in proximity zone size over such short timescales is comparable to the uncertainty on the proximity zone measurement introduced by the redshift estimate, and thus our model is not sensitive to such extreme variability. 

However, despite the lack of sensitivity of quasar proximity zones at $z\sim 6$ to extreme variability and potential previous episodes of quasar activity, there are several other indicators that show that flickering quasar light curves, i.e. multiple epochs of black hole growth, cannot account for the complete discrepancy between the estimated short quasar lifetimes and the accretion times required to grow a billion solar mass black hole from a stellar remnant black hole seed. For instance, the IGM damping wing feature observed in quasars at $z\gtrsim 7$, where the surrounding intergalactic gas still has a high neutral fraction, allows for an integrated constraint on the total number of emitted ionizing photons, i.e. a measurement of the quasars' duty cycle. Such measurements around four known $z\gtrsim 7$ quasars have shown that the total time these quasars have been emitting ionizing radiation over the history of the universe is $t_{\rm Q}\lesssim 10^6$~yr \citep[][Davies et al. in prep.]{Davies2019b}. Thus, the need for modifications to the standard black hole growth model, such as radiatively inefficient accretion or obscured black hole growth phases, remains even in the presence of flickering quasar light curves. 

Additional constraints on flickering quasar light curves and the concomitant growth of SMBHs at $z\sim 6$ can be obtained from the observed distribution of proximity zone sizes. \citet{Davies2019a} model how quasar variability on $\sim 10^5$~yr timescales is imprinted onto the distribution of proximity zone sizes and show that large variations in the ionizing luminosity of quasars on timescales of $\lesssim 10^5$~yr are disfavored based on the good agreement between the bulk of the observed distribution of $R_{\rm p}$ and the model prediction from light--bulb light curves. 

Another argument in favor of light--bulb light curves with $t_{\rm Q}\sim10^6$~yr is based on measurements of the proximity zones in the \ion{He}{2} \lya\ forest. Since the equilibration timescale for \ion{He}{2} is longer than for \ion{H}{1}, i.e. $t_{\rm eq,\,He\,II}\sim 3 \times 10^7$~yr, the proximity zones in the \ion{He}{2} \lya\ forest are sensitive to longer quasar lifetimes and less sensitive to variability on short timescales. Nevertheless, studies of \ion{He}{2} \lya\ proximity zones provide evidence for a lifetime of $t_{\rm Q}\sim10^6$~yr for the quasar population \citep[][]{Khrykin2019, Khrykin2021}. 

Therefore it is highly likely that either radiatively inefficient mass accretion rates, or dust--enshrouded, UV obscured black hole growth phases (or a combination of both) are required to explain the rapid assembly of SMBHs \citep{Eilers2018b, Davies2019b}. 

\section{Summary}\label{sec:summary}

We measure the lifetimes of a sample of ten quasars at $z\sim 6$ based on the extents of their \ion{H}{1} proximity zones observed in the rest-frame UV spectra. This sample was pre--selected out of a parent sample of $122$ quasars to show very small proximity zones and thus likely indicate short quasar lifetimes. In \citetalias{Eilers2020} we obtained sub-mm observations for measurements of the quasars' systemic redshifts, as well as deep optical/NIR spectra to precisely measure the proximity zone sizes for these quasars. 

For four quasars in our sample we estimate extremely short lifetimes, i.e. $\log_{10}(t_{\rm Q}/\rm yr)< 4$, which increases the known young quasar sample to seven at $5.8\lesssim z \lesssim 6.3$ including the previously discovered young objects \citep[]{Eilers2017a, Eilers2018b, Davies2020, Andika2020}. These young quasars contribute $5\%$--$10\%$ of the quasar population at large \citepalias{Eilers2020}. For the remaining six objects in our sample we measure longer quasar lifetimes of $t_{\rm Q}\gtrsim 10^5$~yr. Our results are consistent with an \textit{average} effective quasar lifetime of $t_{\rm Q}\sim 10^6$~yr for an unbiased, i.e. not pre--selected towards young ages, ensemble of quasars \citep[e.g.][Morey et al. subm.]{Davies2019b, Khrykin2016, Khrykin2021}, for which one would expect to find newly turned on quasars $\sim1-10\%$ of the time.  

\subsection{Future Prospects}

In order to explain the rapid growth of SMBHs in the host galaxies of high--redshift quasars, our results provide evidence for radiatively inefficient ($\epsilon \ll 0.1$) mass accretion rates or highly UV obscured, dust--enshrouded black hole growth phases. In either scenario the onset of the UV luminous quasar phase in the discovered young objects happened only very recently, i.e. $\lesssim 10^4$~yr before the time of observations, which implies that whatever process triggered the nuclear activity in these objects might still be observable. Thus the young quasar population represents unique targets to study possible triggering and feedback mechanisms of SMBHs. Future sub-mm observations with high spatial resolution might expose traces of a recent merger or an interaction with a companion galaxy that could have just triggered the nuclear activity in these quasars, or reveal diffuse dust from a recent blow-out of enveloping gas and dust layers. 

Further progress to distinguish between the different scenarios to explain the growth of SMBHs can be made by combining different techniques to estimate quasar lifetimes. For instance, Integral Field Unit (IFU) observations of the extended emission around quasars with very small proximity zones, could provide an additional estimate on the quasars' lifetimes based on the ionization echo of the emission. If the quasars' ionizing radiation has been obscured by dust along our line-of-sight, but the quasars have been accreting matter onto the SMBHs for a much longer time, the ionized regions around the quasars are expected to be extended along unobscured sightlines. On the other hand the extended ionized nebulae are expected to be small or not at all present if the quasar's radiation has indeed just turned on recently and radiatively inefficient accretion rates might explain the rapid growth of SMBHs \citep[see][for details]{Eilers2018b}. Thus, future IFU observations of \lya\ halos or \oiii\ nebulae around young quasars observed for instance with the Multi Unit Spectroscopic Explorer (MUSE) on the VLT, NIRSpec IFU on the James Webb Space Telescope (JWST) or the upcoming Large Lenslet Array Magellan Spectrograph (LLAMAS) on the Magellan telescopes, will enable us to gain new insights into the timescales of quasar activity and the formation and growth of SMBHs in the early universe. 

\acknowledgements

ACE acknowledges support by NASA through the NASA Hubble Fellowship grant $\#$HF2-51434 awarded by the Space Telescope Science Institute, which is operated by the Association of Universities for Research in Astronomy, Inc., for NASA, under contract NAS5-26555. 

\software{numpy \citep{numpy}, scipy \citep{scipy}, matplotlib \citep{matplotlib}, astropy \citep{astropy}}

\appendix

\section{Influence of the Quasars' Spectral Index}\label{abs1}

\begin{figure*}[!h]
\centering
\includegraphics[width=\textwidth]{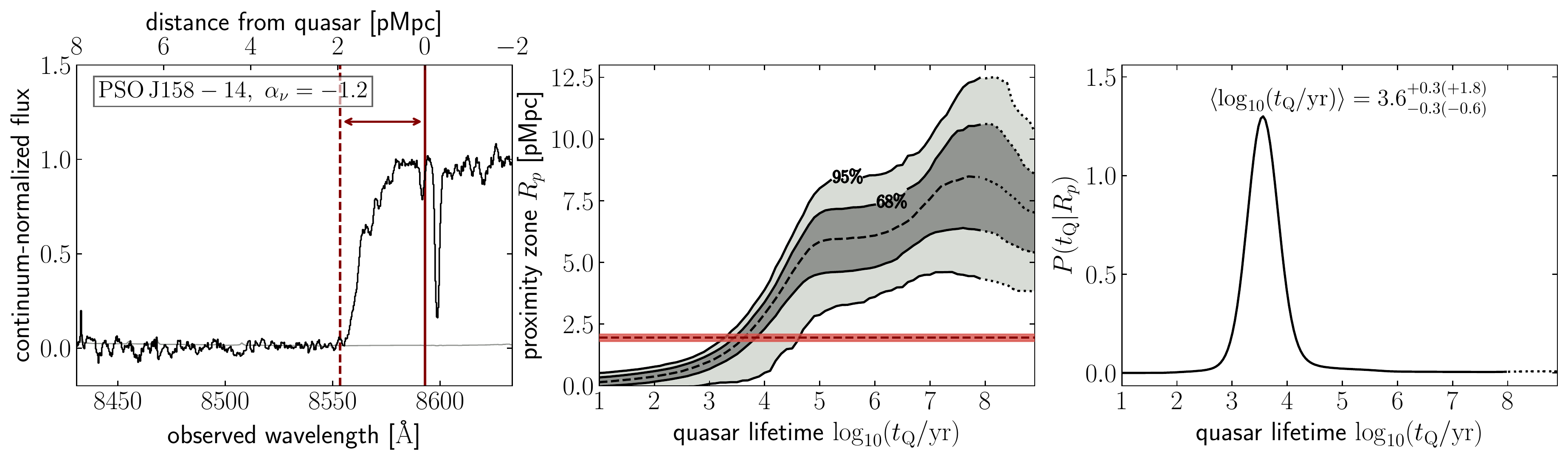}
\includegraphics[width=\textwidth]{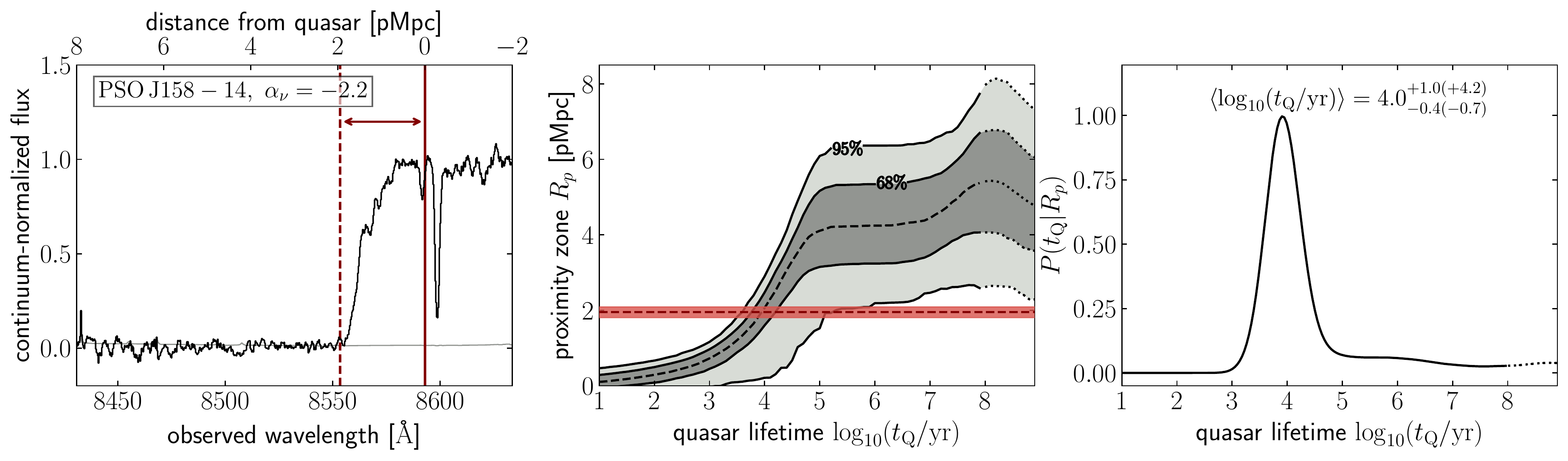}
\caption{Same as the second row of panels in Fig.~\ref{fig:young_qsos}, showing the effects of changing the spectral index of the SED to $\alpha_{\nu}=-1.2$ (\textit{top}) and $\alpha_{\nu}=-2.2$ (\textit{bottom}) for one exemplary young quasar PSO\,J158-14. } \label{fig:SED}
\end{figure*}

The choice of SED alters the number of emitted ionizing photons. As discussed in \S~\ref{sec:caveats} our model assumes a spectral index of $\alpha_{\nu}=-1.7$ determined by \citet{Lusso2015}. In Fig.~\ref{fig:SED} we show the effects of the choice of spectral index by $\pm0.5$ on our results on the quasar lifetime. 

\bibliography{literatur_hz}

\begin{thebibliography}{}
\expandafter\ifx\csname natexlab\endcsname\relax\def\natexlab#1{#1}\fi
\providecommand{\url}[1]{\href{#1}{#1}}
\providecommand{\dodoi}[1]{doi:~\href{http://doi.org/#1}{\nolinkurl{#1}}}
\providecommand{\doeprint}[1]{\href{http://ascl.net/#1}{\nolinkurl{http://ascl.net/#1}}}
\providecommand{\doarXiv}[1]{\href{https://arxiv.org/abs/#1}{\nolinkurl{https://arxiv.org/abs/#1}}}

\bibitem[{{Adelberger}(2004)}]{Adelberger2004}
{Adelberger}, K.~L. 2004, \apj, 612, 706, \dodoi{10.1086/422804}

\bibitem[{{Almgren} {et~al.}(2013){Almgren}, {Bell}, {Lijewski}, {Luki{\'c}},
  \& {Van Andel}}]{Almgren2013}
{Almgren}, A.~S., {Bell}, J.~B., {Lijewski}, M.~J., {Luki{\'c}}, Z., \& {Van
  Andel}, E. 2013, Astrophysical Journal, 765, 39,
  \dodoi{10.1088/0004-637X/765/1/39}

\bibitem[{{Arrigoni Battaia} {et~al.}(2019){Arrigoni Battaia}, {Hennawi},
  {Prochaska}, {O{\~n}orbe}, {Farina}, {Cantalupo}, \& {Lusso}}]{FAB2019}
{Arrigoni Battaia}, F., {Hennawi}, J.~F., {Prochaska}, J.~X., {et~al.} 2019,
  \mnras, 482, 3162, \dodoi{10.1093/mnras/sty2827}

\bibitem[{{Ba{\~n}ados} {et~al.}(2018){Ba{\~n}ados}, {Venemans},
  {Mazzucchelli}, {Farina}, {Walter}, {Wang}, {Decarli}, {Stern}, {Fan},
  {Davies}, {Hennawi}, {Simcoe}, {Turner}, {Rix}, {Yang}, {Kelson}, {Rudie}, \&
  {Winters}}]{Banados2018}
{Ba{\~n}ados}, E., {Venemans}, B.~P., {Mazzucchelli}, C., {et~al.} 2018, \nat,
  553, 473, \dodoi{10.1038/nature25180}

\bibitem[{{Bajtlik} {et~al.}(1988){Bajtlik}, {Duncan}, \&
  {Ostriker}}]{Bajtlik1988}
{Bajtlik}, S., {Duncan}, R.~C., \& {Ostriker}, J.~P. 1988, The Astrophysical
  Journal, 327, 570, \dodoi{10.1086/166217}

\bibitem[{{Becker} {et~al.}(2015){Becker}, {Bolton}, {Madau}, {Pettini},
  {Ryan-Weber}, \& {Venemans}}]{Becker2015}
{Becker}, G.~D., {Bolton}, J.~S., {Madau}, P., {et~al.} 2015, Monthly Notices
  of the Royal Astronomical Society, 447, 3402, \dodoi{10.1093/mnras/stu2646}

\bibitem[{{Becker} {et~al.}(2019){Becker}, {Pettini}, {Rafelski}, {D'Odorico},
  {Boera}, {Christensen}, {Cupani}, {Ellison}, {Farina}, {Fumagalli},
  {L{\'o}pez}, {Neeleman}, {Ryan-Weber}, \& {Worseck}}]{Becker2019}
{Becker}, G.~D., {Pettini}, M., {Rafelski}, M., {et~al.} 2019, \apj, 883, 163,
  \dodoi{10.3847/1538-4357/ab3eb5}

\bibitem[{{Begelman} \& {Volonteri}(2017)}]{BegelmanVolonteri2017}
{Begelman}, M.~C., \& {Volonteri}, M. 2017, \mnras, 464, 1102,
  \dodoi{10.1093/mnras/stw2446}

\bibitem[{{Boera} {et~al.}(2019){Boera}, {Becker}, {Bolton}, \&
  {Nasir}}]{Boera2019}
{Boera}, E., {Becker}, G.~D., {Bolton}, J.~S., \& {Nasir}, F. 2019, \apj, 872,
  101, \dodoi{10.3847/1538-4357/aafee4}

\bibitem[{{Bolton} {et~al.}(2012){Bolton}, {Becker}, {Raskutti}, {Wyithe},
  {Haehnelt}, \& {Sargent}}]{Bolton2012}
{Bolton}, J.~S., {Becker}, G.~D., {Raskutti}, S., {et~al.} 2012, \mnras, 419,
  2880, \dodoi{10.1111/j.1365-2966.2011.19929.x}

\bibitem[{{Bolton} \& {Haehnelt}(2007)}]{BoltonHaehnelt2007a}
{Bolton}, J.~S., \& {Haehnelt}, M.~G. 2007, Monthly Notices of the Royal
  Astronomical Society, 374, 493, \dodoi{10.1111/j.1365-2966.2006.11176.x}

\bibitem[{{Borisova} {et~al.}(2016{\natexlab{a}}){Borisova}, {Lilly},
  {Cantalupo}, {Prochaska}, {Rakic}, \& {Worseck}}]{Borisova2016}
{Borisova}, E., {Lilly}, S.~J., {Cantalupo}, S., {et~al.} 2016{\natexlab{a}},
  \apj, 830, 120, \dodoi{10.3847/0004-637X/830/2/120}

\bibitem[{{Borisova} {et~al.}(2016{\natexlab{b}}){Borisova}, {Cantalupo},
  {Lilly}, {Marino}, {Gallego}, {Bacon}, {Blaizot}, {Bouch{\'e}}, {Brinchmann},
  {Carollo}, {Caruana}, {Finley}, {Herenz}, {Richard}, {Schaye}, {Straka},
  {Turner}, {Urrutia}, {Verhamme}, \& {Wisotzki}}]{Borsiova2016_nebulae}
{Borisova}, E., {Cantalupo}, S., {Lilly}, S.~J., {et~al.} 2016{\natexlab{b}},
  \apj, 831, 39, \dodoi{10.3847/0004-637X/831/1/39}

\bibitem[{{Bosman} {et~al.}(2018){Bosman}, {Fan}, {Jiang}, {Reed}, {Matsuoka},
  {Becker}, \& {Haehnelt}}]{Bosman2018}
{Bosman}, S.~E.~I., {Fan}, X., {Jiang}, L., {et~al.} 2018, \mnras, 479, 1055,
  \dodoi{10.1093/mnras/sty1344}

\bibitem[{{Bosman} {et~al.}(2020{\natexlab{a}}){Bosman}, {Kakiichi}, {Meyer},
  {Gronke}, {Laporte}, \& {Ellis}}]{Bosman2019}
{Bosman}, S. E.~I., {Kakiichi}, K., {Meyer}, R.~A., {et~al.}
  2020{\natexlab{a}}, \apj, 896, 49, \dodoi{10.3847/1538-4357/ab85cd}

\bibitem[{{Bosman} {et~al.}(2020{\natexlab{b}}){Bosman},
  {{\v{D}}urov{\v{c}}{\'\i}kov{\'a}}, {Davies}, \& {Eilers}}]{Bosman2020}
{Bosman}, S.~E.~I., {{\v{D}}urov{\v{c}}{\'\i}kov{\'a}}, D., {Davies}, F.~B., \&
  {Eilers}, A.~C. 2020{\natexlab{b}}, arXiv e-prints, arXiv:2006.10744.
\newblock \doarXiv{2006.10744}

\bibitem[{{Cantalupo} {et~al.}(2014){Cantalupo}, {Arrigoni-Battaia},
  {Prochaska}, {Hennawi}, \& {Madau}}]{Cantalupo2014}
{Cantalupo}, S., {Arrigoni-Battaia}, F., {Prochaska}, J.~X., {Hennawi}, J.~F.,
  \& {Madau}, P. 2014, \nat, 506, 63, \dodoi{10.1038/nature12898}

\bibitem[{{Carilli} {et~al.}(2010){Carilli}, {Wang}, {Fan}, {Walter}, {Kurk},
  {Riechers}, {Wagg}, {Hennawi}, {Jiang}, {Menten}, {Bertoldi}, {Strauss}, \&
  {Cox}}]{Carilli2010}
{Carilli}, C.~L., {Wang}, R., {Fan}, X., {et~al.} 2010, \apj, 714, 834,
  \dodoi{10.1088/0004-637X/714/1/834}

\bibitem[{{Chen} \& {Gnedin}(2018)}]{Chen2018}
{Chen}, H., \& {Gnedin}, N.~Y. 2018, \apj, 868, 126,
  \dodoi{10.3847/1538-4357/aae8e8}

\bibitem[{{Chen} \& {Gnedin}(2020)}]{ChenGnedin2020}
---. 2020, arXiv e-prints, arXiv:2008.04911.
\newblock \doarXiv{2008.04911}

\bibitem[{{Cieplak} \& {Slosar}(2017)}]{Cieplak2017}
{Cieplak}, A.~M., \& {Slosar}, A. 2017, \jcap, 2017, 013,
  \dodoi{10.1088/1475-7516/2017/10/013}

\bibitem[{{Conroy} \& {White}(2013)}]{ConroyWhite2013}
{Conroy}, C., \& {White}, M. 2013, The Astrophysical Journal, 762, 70,
  \dodoi{10.1088/0004-637X/762/2/70}

\bibitem[{{Cooper} {et~al.}(2019){Cooper}, {Simcoe}, {Cooksey}, {Bordoloi},
  {Miller}, {Furesz}, {Turner}, \& {Ba{\~n}ados}}]{Cooper2019}
{Cooper}, T.~J., {Simcoe}, R.~A., {Cooksey}, K.~L., {et~al.} 2019, \apj, 882,
  77, \dodoi{10.3847/1538-4357/ab3402}

\bibitem[{{Davies}(2020)}]{Davies2020_ghost}
{Davies}, F.~B. 2020, \mnras, 494, 2937, \dodoi{10.1093/mnras/staa528}

\bibitem[{{Davies} {et~al.}(2018{\natexlab{a}}){Davies}, {Becker}, \&
  {Furlanetto}}]{Davies2018b}
{Davies}, F.~B., {Becker}, G.~D., \& {Furlanetto}, S.~R. 2018{\natexlab{a}},
  \apj, 860, 155, \dodoi{10.3847/1538-4357/aac2d6}

\bibitem[{{Davies} {et~al.}(2016){Davies}, {Furlanetto}, \&
  {McQuinn}}]{Davies2016}
{Davies}, F.~B., {Furlanetto}, S.~R., \& {McQuinn}, M. 2016, \mnras, 457, 3006,
  \dodoi{10.1093/mnras/stw055}

\bibitem[{{Davies} {et~al.}(2019){Davies}, {Hennawi}, \&
  {Eilers}}]{Davies2019b}
{Davies}, F.~B., {Hennawi}, J.~F., \& {Eilers}, A.-C. 2019, \apjl, 884, L19,
  \dodoi{10.3847/2041-8213/ab42e3}

\bibitem[{{Davies} {et~al.}(2020{\natexlab{a}}){Davies}, {Hennawi}, \&
  {Eilers}}]{Davies2019a}
---. 2020{\natexlab{a}}, \mnras, 493, 1330, \dodoi{10.1093/mnras/stz3303}

\bibitem[{{Davies} {et~al.}(2018{\natexlab{b}}){Davies}, {Hennawi}, {Eilers},
  \& {Luki{\'c}}}]{Davies2018a}
{Davies}, F.~B., {Hennawi}, J.~F., {Eilers}, A.-C., \& {Luki{\'c}}, Z.
  2018{\natexlab{b}}, \apj, 855, 106, \dodoi{10.3847/1538-4357/aaaf70}

\bibitem[{{Davies} {et~al.}(2020{\natexlab{b}}){Davies}, {Wang}, {Eilers}, \&
  {Hennawi}}]{Davies2020}
{Davies}, F.~B., {Wang}, F., {Eilers}, A.-C., \& {Hennawi}, J.~F.
  2020{\natexlab{b}}, arXiv e-prints, arXiv:2007.15657.
\newblock \doarXiv{2007.15657}

\bibitem[{{Davies} {et~al.}(2018{\natexlab{c}}){Davies}, {Hennawi},
  {Ba{\~n}ados}, {Simcoe}, {Decarli}, {Fan}, {Farina}, {Mazzucchelli}, {Rix},
  {Venemans}, {Walter}, {Wang}, \& {Yang}}]{Davies2018}
{Davies}, F.~B., {Hennawi}, J.~F., {Ba{\~n}ados}, E., {et~al.}
  2018{\natexlab{c}}, \apj, 864, 143, \dodoi{10.3847/1538-4357/aad7f8}

\bibitem[{{Decarli} {et~al.}(2018){Decarli}, {Walter}, {Venemans},
  {Ba{\~n}ados}, {Bertoldi}, {Carilli}, {Fan}, {Farina}, {Mazzucchelli},
  {Riechers}, {Rix}, {Strauss}, {Wang}, \& {Yang}}]{Decarli2018}
{Decarli}, R., {Walter}, F., {Venemans}, B.~P., {et~al.} 2018, \apj, 854, 97,
  \dodoi{10.3847/1538-4357/aaa5aa}

\bibitem[{{Di Matteo} {et~al.}(2005){Di Matteo}, {Springel}, \&
  {Hernquist}}]{DiMatteo2005}
{Di Matteo}, T., {Springel}, V., \& {Hernquist}, L. 2005, \nat, 433, 604,
  \dodoi{10.1038/nature03335}

\bibitem[{{D'Odorico} {et~al.}(2013){D'Odorico}, {Cupani}, {Cristiani},
  {Maiolino}, {Molaro}, {Nonino}, {Centuri{\'o}n}, {Cimatti}, {di Serego
  Alighieri}, {Fiore}, {Fontana}, {Gallerani}, {Giallongo}, {Mannucci},
  {Marconi}, {Pentericci}, {Viel}, \& {Vladilo}}]{Dodorico2013}
{D'Odorico}, V., {Cupani}, G., {Cristiani}, S., {et~al.} 2013, \mnras, 435,
  1198, \dodoi{10.1093/mnras/stt1365}

\bibitem[{{Drake} {et~al.}(2019){Drake}, {Farina}, {Neeleman}, {Walter},
  {Venemans}, {Banados}, {Mazzucchelli}, \& {Decarli}}]{Drake2019}
{Drake}, A.~B., {Farina}, E.~P., {Neeleman}, M., {et~al.} 2019, \apj, 881, 131,
  \dodoi{10.3847/1538-4357/ab2984}

\bibitem[{{Efstathiou} \& {Rees}(1988)}]{EfstathiouRees1988}
{Efstathiou}, G., \& {Rees}, M.~J. 1988, \mnras, 230, 5p,
  \dodoi{10.1093/mnras/230.1.5P}

\bibitem[{{Eftekharzadeh} {et~al.}(2015){Eftekharzadeh}, {Myers}, {White},
  {Weinberg}, {Schneider}, {Shen}, {Font-Ribera}, {Ross}, {Paris}, \&
  {Streblyanska}}]{Eftekharzadeh2015}
{Eftekharzadeh}, S., {Myers}, A.~D., {White}, M., {et~al.} 2015, \mnras, 453,
  2779, \dodoi{10.1093/mnras/stv1763}

\bibitem[{{Eilers} {et~al.}(2018{\natexlab{a}}){Eilers}, {Davies}, \&
  {Hennawi}}]{Eilers2018a}
{Eilers}, A.-C., {Davies}, F.~B., \& {Hennawi}, J.~F. 2018{\natexlab{a}}, \apj,
  864, 53, \dodoi{10.3847/1538-4357/aad4fd}

\bibitem[{{Eilers} {et~al.}(2017{\natexlab{a}}){Eilers}, {Davies}, {Hennawi},
  {Prochaska}, {Luki{\'c}}, \& {Mazzucchelli}}]{Eilers2017a}
{Eilers}, A.-C., {Davies}, F.~B., {Hennawi}, J.~F., {et~al.}
  2017{\natexlab{a}}, \apj, 840, 24, \dodoi{10.3847/1538-4357/aa6c60}

\bibitem[{{Eilers} {et~al.}(2018{\natexlab{b}}){Eilers}, {Hennawi}, \&
  {Davies}}]{Eilers2018b}
{Eilers}, A.-C., {Hennawi}, J.~F., \& {Davies}, F.~B. 2018{\natexlab{b}}, \apj,
  867, 30, \dodoi{10.3847/1538-4357/aae081}

\bibitem[{{Eilers} {et~al.}(2019){Eilers}, {Hennawi}, {Davies}, \&
  {O{\~n}orbe}}]{Eilers2019b}
{Eilers}, A.-C., {Hennawi}, J.~F., {Davies}, F.~B., \& {O{\~n}orbe}, J. 2019,
  \apj, 881, 23, \dodoi{10.3847/1538-4357/ab2b3f}

\bibitem[{{Eilers} {et~al.}(2017{\natexlab{b}}){Eilers}, {Hennawi}, \&
  {Lee}}]{Eilers2017b}
{Eilers}, A.-C., {Hennawi}, J.~F., \& {Lee}, K.-G. 2017{\natexlab{b}}, \apj,
  844, 136, \dodoi{10.3847/1538-4357/aa7e31}

\bibitem[{{Eilers} {et~al.}(2020){Eilers}, {Hennawi}, {Decarli}, {Davies},
  {Venemans}, {Walter}, {Ba{\~n}ados}, {Fan}, {Farina}, {Mazzucchelli},
  {Novak}, {Schindler}, {Simcoe}, {Wang}, \& {Yang}}]{Eilers2020}
{Eilers}, A.-C., {Hennawi}, J.~F., {Decarli}, R., {et~al.} 2020, \apj, 900, 37,
  \dodoi{10.3847/1538-4357/aba52e}

\bibitem[{{Fan} {et~al.}(2000){Fan}, {White}, {Davis}, {Becker}, {Strauss},
  {Haiman}, {Schneider}, {Gregg}, {Gunn}, {Knapp}, {Lupton}, {Anderson},
  {Anderson}, {Annis}, {Bahcall}, {Boroski}, {Brunner}, {Chen}, {Connolly},
  {Csabai}, {Doi}, {Fukugita}, {Hennessy}, {Hindsley}, {Ichikawa},
  {Ivezi{\'c}}, {Loveday}, {Meiksin}, {McKay}, {Munn}, {Newberg}, {Nichol},
  {Okamura}, {Pier}, {Sekiguchi}, {Shimasaku}, {Stoughton}, {Szalay},
  {Szokoly}, {Thakar}, {Vogeley}, \& {York}}]{Fan2000}
{Fan}, X., {White}, R.~L., {Davis}, M., {et~al.} 2000, \aj, 120, 1167,
  \dodoi{10.1086/301534}

\bibitem[{{Fan} {et~al.}(2001){Fan}, {Narayanan}, {Lupton}, {Strauss}, {Knapp},
  {Becker}, {White}, {Pentericci}, {Leggett}, {Haiman}, {Gunn}, {Ivezi{\'c}},
  {Schneider}, {Anderson}, {Brinkmann}, {Bahcall}, {Connolly}, {Csabai}, {Doi},
  {Fukugita}, {Geballe}, {Grebel}, {Harbeck}, {Hennessy}, {Lamb}, {Miknaitis},
  {Munn}, {Nichol}, {Okamura}, {Pier}, {Prada}, {Richards}, {Szalay}, \&
  {York}}]{Fan2001}
{Fan}, X., {Narayanan}, V.~K., {Lupton}, R.~H., {et~al.} 2001, \aj, 122, 2833,
  \dodoi{10.1086/324111}

\bibitem[{{Fan} {et~al.}(2006){Fan}, {Strauss}, {Becker}, {White}, {Gunn},
  {Knapp}, {Richards}, {Schneider}, {Brinkmann}, \& {Fukugita}}]{Fan2006}
{Fan}, X., {Strauss}, M.~A., {Becker}, R.~H., {et~al.} 2006, The Astronomical
  Journal, 132, 117, \dodoi{10.1086/504836}

\bibitem[{{Farina} {et~al.}(2019){Farina}, {Arrigoni-Battaia}, {Costa},
  {Walter}, {Hennawi}, {Drake}, {Decarli}, {Gutcke}, {Mazzucchelli},
  {Neeleman}, {Georgiev}, {Eilers}, {Davies}, {Ba{\~n}ados}, {Fan}, {Onoue},
  {Schindler}, {Venemans}, {Wang}, {Yang}, {Rabien}, \& {Busoni}}]{Farina2019}
{Farina}, E.~P., {Arrigoni-Battaia}, F., {Costa}, T., {et~al.} 2019, \apj, 887,
  196, \dodoi{10.3847/1538-4357/ab5847}

\bibitem[{{Fujimoto} {et~al.}(2020){Fujimoto}, {Oguri}, {Nagao}, {Izumi}, \&
  {Ouchi}}]{Fujimoto2020}
{Fujimoto}, S., {Oguri}, M., {Nagao}, T., {Izumi}, T., \& {Ouchi}, M. 2020,
  \apj, 891, 64, \dodoi{10.3847/1538-4357/ab718c}

\bibitem[{{Fumagalli} {et~al.}(2013){Fumagalli}, {O'Meara}, {Prochaska}, \&
  {Worseck}}]{Fumagalli2013}
{Fumagalli}, M., {O'Meara}, J.~M., {Prochaska}, J.~X., \& {Worseck}, G. 2013,
  \apj, 775, 78, \dodoi{10.1088/0004-637X/775/1/78}

\bibitem[{{Greig} {et~al.}(2017){Greig}, {Mesinger}, {Haiman}, \&
  {Simcoe}}]{Greig2017}
{Greig}, B., {Mesinger}, A., {Haiman}, Z., \& {Simcoe}, R.~A. 2017, \mnras,
  466, 4239, \dodoi{10.1093/mnras/stw3351}

\bibitem[{{Haiman} \& {Hui}(2001)}]{HaimanHui2001}
{Haiman}, Z., \& {Hui}, L. 2001, The Astrophysical Journal, 547, 27,
  \dodoi{10.1086/318330}

\bibitem[{{Hennawi} \& {Prochaska}(2013)}]{Hennawi2013}
{Hennawi}, J.~F., \& {Prochaska}, J.~X. 2013, \apj, 766, 58,
  \dodoi{10.1088/0004-637X/766/1/58}

\bibitem[{{Hennawi} {et~al.}(2015){Hennawi}, {Prochaska}, {Cantalupo}, \&
  {Arrigoni-Battaia}}]{Hennawi2015}
{Hennawi}, J.~F., {Prochaska}, J.~X., {Cantalupo}, S., \& {Arrigoni-Battaia},
  F. 2015, Science, 348, 779, \dodoi{10.1126/science.aaa5397}

\bibitem[{{Hennawi} {et~al.}(2006){Hennawi}, {Prochaska}, {Burles}, {Strauss},
  {Richards}, {Schlegel}, {Fan}, {Schneider}, {Zakamska}, {Oguri}, {Gunn},
  {Lupton}, \& {Brinkmann}}]{Hennawi2006}
{Hennawi}, J.~F., {Prochaska}, J.~X., {Burles}, S., {et~al.} 2006, \apj, 651,
  61, \dodoi{10.1086/507069}

\bibitem[{{Hogan} {et~al.}(1997){Hogan}, {Anderson}, \& {Rugers}}]{Hogan1997}
{Hogan}, C.~J., {Anderson}, S.~F., \& {Rugers}, M.~H. 1997, \aj, 113, 1495,
  \dodoi{10.1086/118366}

\bibitem[{{Hopkins} {et~al.}(2005{\natexlab{a}}){Hopkins}, {Hernquist}, {Cox},
  {Di Matteo}, {Martini}, {Robertson}, \& {Springel}}]{Hopkins2005a}
{Hopkins}, P.~F., {Hernquist}, L., {Cox}, T.~J., {et~al.} 2005{\natexlab{a}},
  \apj, 630, 705, \dodoi{10.1086/432438}

\bibitem[{{Hopkins} {et~al.}(2008){Hopkins}, {Hernquist}, {Cox}, \&
  {Kere{\v{s}}}}]{Hopkins2008}
{Hopkins}, P.~F., {Hernquist}, L., {Cox}, T.~J., \& {Kere{\v{s}}}, D. 2008,
  \apjs, 175, 356, \dodoi{10.1086/524362}

\bibitem[{{Hopkins} {et~al.}(2005{\natexlab{b}}){Hopkins}, {Hernquist},
  {Martini}, {Cox}, {Robertson}, {Di Matteo}, \& {Springel}}]{Hopkins2005}
{Hopkins}, P.~F., {Hernquist}, L., {Martini}, P., {et~al.} 2005{\natexlab{b}},
  \apjl, 625, L71, \dodoi{10.1086/431146}

\bibitem[{Hunter(2007)}]{matplotlib}
Hunter, J.~D. 2007, Computing In Science \& Engineering, 9, 90,
  \dodoi{10.1109/MCSE.2007.55}

\bibitem[{{Inayoshi} {et~al.}(2016){Inayoshi}, {Haiman}, \&
  {Ostriker}}]{Inayoshi2016}
{Inayoshi}, K., {Haiman}, Z., \& {Ostriker}, J.~P. 2016, \mnras, 459, 3738,
  \dodoi{10.1093/mnras/stw836}

\bibitem[{{Ir{\v{s}}i{\v{c}}} {et~al.}(2017){Ir{\v{s}}i{\v{c}}}, {Viel},
  {Haehnelt}, {Bolton}, {Cristiani}, {Becker}, {D'Odorico}, {Cupani}, {Kim},
  {Berg}, {L{\'o}pez}, {Ellison}, {Christensen}, {Denney}, \&
  {Worseck}}]{Irsic2017a}
{Ir{\v{s}}i{\v{c}}}, V., {Viel}, M., {Haehnelt}, M.~G., {et~al.} 2017, \prd,
  96, 023522, \dodoi{10.1103/PhysRevD.96.023522}

\bibitem[{{Ishimoto} {et~al.}(2020){Ishimoto}, {Kashikawa}, {Onoue},
  {Matsuoka}, {Izumi}, {Strauss}, {Fujimoto}, {Imanishi}, {Ito}, {Iwasawa},
  {Kawaguchi}, {Lee}, {Liang}, {Lu}, {Momose}, {Toba}, \&
  {Uchiyama}}]{Ishimoto2020}
{Ishimoto}, R., {Kashikawa}, N., {Onoue}, M., {et~al.} 2020, \apj, 903, 60,
  \dodoi{10.3847/1538-4357/abb80b}

\bibitem[{{Jakobsen} {et~al.}(2003){Jakobsen}, {Jansen}, {Wagner}, \&
  {Reimers}}]{Jakobsen2003}
{Jakobsen}, P., {Jansen}, R.~A., {Wagner}, S., \& {Reimers}, D. 2003, \aap,
  397, 891, \dodoi{10.1051/0004-6361:20021579}

\bibitem[{Jones {et~al.}(2001)Jones, Oliphant, Peterson, {et~al.}}]{scipy}
Jones, E., Oliphant, T., Peterson, P., {et~al.} 2001, {SciPy}: Open source
  scientific tools for {Python}.
\newblock \url{http://www.scipy.org/}

\bibitem[{{Kaiser}(1984)}]{Kaiser1984}
{Kaiser}, N. 1984, \apjl, 284, L9, \dodoi{10.1086/184341}

\bibitem[{{Keating} {et~al.}(2015){Keating}, {Haehnelt}, {Cantalupo}, \&
  {Puchwein}}]{Keating2015}
{Keating}, L.~C., {Haehnelt}, M.~G., {Cantalupo}, S., \& {Puchwein}, E. 2015,
  Monthly Notices of the Royal Astronomical Society, 454, 681,
  \dodoi{10.1093/mnras/stv2020}

\bibitem[{{Keating} {et~al.}(2020){Keating}, {Kulkarni}, {Haehnelt}, {Chardin},
  \& {Aubert}}]{Keating2020}
{Keating}, L.~C., {Kulkarni}, G., {Haehnelt}, M.~G., {Chardin}, J., \&
  {Aubert}, D. 2020, \mnras, 497, 906, \dodoi{10.1093/mnras/staa1909}

\bibitem[{{Keel} {et~al.}(2012){Keel}, {Chojnowski}, {Bennert}, {Schawinski},
  {Lintott}, {Lynn}, {Pancoast}, {Harris}, {Nierenberg}, {Sonnenfeld}, \&
  {Proctor}}]{Keel2012}
{Keel}, W.~C., {Chojnowski}, S.~D., {Bennert}, V.~N., {et~al.} 2012, \mnras,
  420, 878, \dodoi{10.1111/j.1365-2966.2011.20101.x}

\bibitem[{{Khrykin} {et~al.}(2017){Khrykin}, {Hennawi}, \&
  {McQuinn}}]{Khrykin2017}
{Khrykin}, I.~S., {Hennawi}, J.~F., \& {McQuinn}, M. 2017, \apj, 838, 96,
  \dodoi{10.3847/1538-4357/aa6621}

\bibitem[{{Khrykin} {et~al.}(2016){Khrykin}, {Hennawi}, {McQuinn}, \&
  {Worseck}}]{Khrykin2016}
{Khrykin}, I.~S., {Hennawi}, J.~F., {McQuinn}, M., \& {Worseck}, G. 2016, \apj,
  824, 133, \dodoi{10.3847/0004-637X/824/2/133}

\bibitem[{{Khrykin} {et~al.}(2019){Khrykin}, {Hennawi}, \&
  {Worseck}}]{Khrykin2019}
{Khrykin}, I.~S., {Hennawi}, J.~F., \& {Worseck}, G. 2019, \mnras, 484, 3897,
  \dodoi{10.1093/mnras/stz135}

\bibitem[{{Khrykin} {et~al.}(2021){Khrykin}, {Hennawi}, {Worseck}, \&
  {Davies}}]{Khrykin2021}
{Khrykin}, I.~S., {Hennawi}, J.~F., {Worseck}, G., \& {Davies}, F.~B. 2021,
  arXiv e-prints, arXiv:2102.04477.
\newblock \doarXiv{2102.04477}

\bibitem[{{Kirkman} \& {Tytler}(2008)}]{KirkmanTytler2008}
{Kirkman}, D., \& {Tytler}, D. 2008, \mnras, 391, 1457,
  \dodoi{10.1111/j.1365-2966.2008.13994.x}

\bibitem[{{Kulkarni} {et~al.}(2019){Kulkarni}, {Keating}, {Haehnelt}, {Bosman},
  {Puchwein}, {Chardin}, \& {Aubert}}]{Kulkarni2019}
{Kulkarni}, G., {Keating}, L.~C., {Haehnelt}, M.~G., {et~al.} 2019, \mnras,
  485, L24, \dodoi{10.1093/mnrasl/slz025}

\bibitem[{{Laurent} {et~al.}(2017){Laurent}, {Eftekharzadeh}, {Le Goff},
  {Myers}, {Burtin}, {White}, {Ross}, {Tinker}, {Tojeiro}, {Bautista},
  {Brinkmann}, {Comparat}, {Dawson}, {du Mas des Bourboux}, {Kneib}, {McGreer},
  {Palanque-Delabrouille}, {Percival}, {Prada}, {Rossi}, {Schneider},
  {Weinberg}, {Y{\`e}che}, {Zarrouk}, \& {Zhao}}]{Laurent2017}
{Laurent}, P., {Eftekharzadeh}, S., {Le Goff}, J.-M., {et~al.} 2017, \jcap,
  2017, 017, \dodoi{10.1088/1475-7516/2017/07/017}

\bibitem[{{Lee} {et~al.}(2015){Lee}, {Hennawi}, {Spergel}, {Weinberg}, {Hogg},
  {Viel}, {Bolton}, {Bailey}, {Pieri}, {Carithers}, {Schlegel}, {Lundgren},
  {Palanque-Delabrouille}, {Suzuki}, {Schneider}, \& {Y{\`e}che}}]{Lee2015}
{Lee}, K.-G., {Hennawi}, J.~F., {Spergel}, D.~N., {et~al.} 2015, \apj, 799,
  196, \dodoi{10.1088/0004-637X/799/2/196}

\bibitem[{{Lintott} {et~al.}(2009){Lintott}, {Schawinski}, {Keel}, {van Arkel},
  {Bennert}, {Edmondson}, {Thomas}, {Smith}, {Herbert}, {Jarvis}, {Virani},
  {Andreescu}, {Bamford}, {Land}, {Murray}, {Nichol}, {Raddick}, {Slosar},
  {Szalay}, \& {Vandenberg}}]{Lintott2009}
{Lintott}, C.~J., {Schawinski}, K., {Keel}, W., {et~al.} 2009, \mnras, 399,
  129, \dodoi{10.1111/j.1365-2966.2009.15299.x}

\bibitem[{{Luki{\'c}} {et~al.}(2015){Luki{\'c}}, {Stark}, {Nugent}, {White},
  {Meiksin}, \& {Almgren}}]{Lukic2015}
{Luki{\'c}}, Z., {Stark}, C.~W., {Nugent}, P., {et~al.} 2015, \mnras, 446,
  3697, \dodoi{10.1093/mnras/stu2377}

\bibitem[{{Lusso} {et~al.}(2015){Lusso}, {Worseck}, {Hennawi}, {Prochaska},
  {Vignali}, {Stern}, \& {O'Meara}}]{Lusso2015}
{Lusso}, E., {Worseck}, G., {Hennawi}, J.~F., {et~al.} 2015, Monthly Notices of
  the Royal Astronomical Society, 449, 4204, \dodoi{10.1093/mnras/stv516}

\bibitem[{{Madau} \& {Rees}(2001)}]{MadauRees2001}
{Madau}, P., \& {Rees}, M.~J. 2001, \apjl, 551, L27, \dodoi{10.1086/319848}

\bibitem[{{Martini} \& {Weinberg}(2001)}]{MartiniWeinberg2001}
{Martini}, P., \& {Weinberg}, D.~H. 2001, \apj, 547, 12, \dodoi{10.1086/318331}

\bibitem[{{Mazzucchelli} {et~al.}(2017){Mazzucchelli}, {Ba{\~n}ados},
  {Venemans}, {Decarli}, {Farina}, {Walter}, {Eilers}, {Rix}, {Simcoe},
  {Stern}, {Fan}, {Schlafly}, {De Rosa}, {Hennawi}, {Chambers}, {Greiner},
  {Burgett}, {Draper}, {Kaiser}, {Kudritzki}, {Magnier}, {Metcalfe}, {Waters},
  \& {Wainscoat}}]{Mazzucchelli2017}
{Mazzucchelli}, C., {Ba{\~n}ados}, E., {Venemans}, B.~P., {et~al.} 2017, \apj,
  849, 91, \dodoi{10.3847/1538-4357/aa9185}

\bibitem[{{McQuinn} \& {Upton Sanderbeck}(2016)}]{McQuinn2015}
{McQuinn}, M., \& {Upton Sanderbeck}, P.~R. 2016, \mnras, 456, 47,
  \dodoi{10.1093/mnras/stv2675}

\bibitem[{{McQuinn} \& {Werk}(2018)}]{McQuinnWerk2018}
{McQuinn}, M., \& {Werk}, J.~K. 2018, \apj, 852, 33,
  \dodoi{10.3847/1538-4357/aa9d3f}

\bibitem[{{Meiksin} {et~al.}(2010){Meiksin}, {Tittley}, \&
  {Brown}}]{Meiksin2010}
{Meiksin}, A., {Tittley}, E.~R., \& {Brown}, C.~K. 2010, \mnras, 401, 77,
  \dodoi{10.1111/j.1365-2966.2009.15667.x}

\bibitem[{{Meyer} {et~al.}(2019){Meyer}, {Bosman}, \& {Ellis}}]{Meyer2019}
{Meyer}, R.~A., {Bosman}, S. E.~I., \& {Ellis}, R.~S. 2019, \mnras, 487, 3305,
  \dodoi{10.1093/mnras/stz1504}

\bibitem[{{Miralda-Escud{\'e}}(1998)}]{MiraldaEscude1998}
{Miralda-Escud{\'e}}, J. 1998, \apj, 501, 15, \dodoi{10.1086/305799}

\bibitem[{{Mortlock} {et~al.}(2011){Mortlock}, {Warren}, {Venemans}, {Patel},
  {Hewett}, {McMahon}, {Simpson}, {Theuns}, {Gonz{\'a}les-Solares}, {Adamson},
  {Dye}, {Hambly}, {Hirst}, {Irwin}, {Kuiper}, {Lawrence}, \&
  {R{\"o}ttgering}}]{Mortlock2011}
{Mortlock}, D.~J., {Warren}, S.~J., {Venemans}, B.~P., {et~al.} 2011, Nature,
  474, 616, \dodoi{10.1038/nature10159}

\bibitem[{{Neeleman} {et~al.}(2019){Neeleman}, {Ba{\~n}ados}, {Walter},
  {Decarli}, {Venemans}, {Carilli}, {Fan}, {Farina}, {Mazzucchelli}, {Novak},
  {Riechers}, {Rix}, \& {Wang}}]{Neeleman2019}
{Neeleman}, M., {Ba{\~n}ados}, E., {Walter}, F., {et~al.} 2019, \apj, 882, 10,
  \dodoi{10.3847/1538-4357/ab2ed3}

\bibitem[{{Novak} {et~al.}(2011){Novak}, {Ostriker}, \& {Ciotti}}]{Novak2011}
{Novak}, G.~S., {Ostriker}, J.~P., \& {Ciotti}, L. 2011, \apj, 737, 26,
  \dodoi{10.1088/0004-637X/737/1/26}

\bibitem[{{Oppenheimer} {et~al.}(2018){Oppenheimer}, {Segers}, {Schaye},
  {Richings}, \& {Crain}}]{Oppenheimer2018}
{Oppenheimer}, B.~D., {Segers}, M., {Schaye}, J., {Richings}, A.~J., \&
  {Crain}, R.~A. 2018, \mnras, 474, 4740, \dodoi{10.1093/mnras/stx2967}

\bibitem[{{Palanque-Delabrouille} {et~al.}(2013){Palanque-Delabrouille},
  {Y{\`e}che}, {Borde}, {Le Goff}, {Rossi}, {Viel}, {Aubourg}, {Bailey},
  {Bautista}, {Blomqvist}, {Bolton}, {Bolton}, {Busca}, {Carithers}, {Croft},
  {Dawson}, {Delubac}, {Font-Ribera}, {Ho}, {Kirkby}, {Lee}, {Margala},
  {Miralda-Escud{\'e}}, {Muna}, {Myers}, {Noterdaeme}, {P{\^a}ris},
  {Petitjean}, {Pieri}, {Rich}, {Rollinde}, {Ross}, {Schlegel}, {Schneider},
  {Slosar}, \& {Weinberg}}]{Palanque2013}
{Palanque-Delabrouille}, N., {Y{\`e}che}, C., {Borde}, A., {et~al.} 2013, \aap,
  559, A85, \dodoi{10.1051/0004-6361/201322130}

\bibitem[{{P{\^a}ris} {et~al.}(2011){P{\^a}ris}, {Petitjean}, {Rollinde},
  {Aubourg}, {Busca}, {Charlassier}, {Delubac}, {Hamilton}, {Le Goff},
  {Palanque-Delabrouille}, {Peirani}, {Pichon}, {Rich}, {Vargas-Maga{\~n}a}, \&
  {Y{\`e}che}}]{Paris2011}
{P{\^a}ris}, I., {Petitjean}, P., {Rollinde}, E., {et~al.} 2011, Astronomy \&
  Astrophysics, 530, A50, \dodoi{10.1051/0004-6361/201016233}

\bibitem[{{Prochaska} {et~al.}(2020){Prochaska}, {Hennawi}, {Westfall},
  {Cooke}, {Wang}, {Hsyu}, {Davies}, \& {Farina}}]{Prochaska2020}
{Prochaska}, J.~X., {Hennawi}, J.~F., {Westfall}, K.~B., {et~al.} 2020, arXiv
  e-prints, arXiv:2005.06505.
\newblock \doarXiv{2005.06505}

\bibitem[{{Rahmati} {et~al.}(2013){Rahmati}, {Pawlik}, {Rai{\v{c}}evi{\'c}}, \&
  {Schaye}}]{Rahmati2013}
{Rahmati}, A., {Pawlik}, A.~H., {Rai{\v{c}}evi{\'c}}, M., \& {Schaye}, J. 2013,
  \mnras, 430, 2427, \dodoi{10.1093/mnras/stt066}

\bibitem[{{Richards} {et~al.}(2002){Richards}, {Vanden Berk}, {Reichard},
  {Hall}, {Schneider}, {SubbaRao}, {Thakar}, \& {York}}]{Richards2002}
{Richards}, G.~T., {Vanden Berk}, D.~E., {Reichard}, T.~A., {et~al.} 2002, \aj,
  124, 1, \dodoi{10.1086/341167}

\bibitem[{{Rumbaugh} {et~al.}(2018){Rumbaugh}, {Shen}, {Morganson}, {Liu},
  {Banerji}, {McMahon}, {Abdalla}, {Benoit-L{\'e}vy}, {Bertin}, {Brooks},
  {Buckley-Geer}, {Capozzi}, {Carnero Rosell}, {Carrasco Kind}, {Carretero},
  {Cunha}, {D'Andrea}, {da Costa}, {DePoy}, {Desai}, {Doel}, {Frieman},
  {Garc{\'\i}a-Bellido}, {Gruen}, {Gruendl}, {Gschwend}, {Gutierrez},
  {Honscheid}, {James}, {Kuehn}, {Kuhlmann}, {Kuropatkin}, {Lima}, {Maia},
  {Marshall}, {Martini}, {Menanteau}, {Plazas}, {Reil}, {Roodman}, {Sanchez},
  {Scarpine}, {Schindler}, {Schubnell}, {Sheldon}, {Smith}, {Soares-Santos},
  {Sobreira}, {Suchyta}, {Swanson}, {Walker}, {Wester}, \& {DES
  Collaboration}}]{Rumbaugh2018}
{Rumbaugh}, N., {Shen}, Y., {Morganson}, E., {et~al.} 2018, \apj, 854, 160,
  \dodoi{10.3847/1538-4357/aaa9b6}

\bibitem[{{Runnoe} {et~al.}(2012){Runnoe}, {Brotherton}, \&
  {Shang}}]{Runnoe2012}
{Runnoe}, J.~C., {Brotherton}, M.~S., \& {Shang}, Z. 2012, \mnras, 422, 478,
  \dodoi{10.1111/j.1365-2966.2012.20620.x}

\bibitem[{{Salpeter}(1964)}]{Salpeter1964}
{Salpeter}, E.~E. 1964, The Astrophysical Journal, 140, 796,
  \dodoi{10.1086/147973}

\bibitem[{{Sartori} {et~al.}(2018){Sartori}, {Schawinski}, {Koss}, {Ricci},
  {Treister}, {Stern}, {Lansbury}, {Maksym}, {Balokovi{\'c}}, {Gandhi}, {Keel},
  \& {Ballantyne}}]{Sartori2018}
{Sartori}, L.~F., {Schawinski}, K., {Koss}, M.~J., {et~al.} 2018, \mnras, 474,
  2444, \dodoi{10.1093/mnras/stx2952}

\bibitem[{{Schawinski} {et~al.}(2015){Schawinski}, {Koss}, {Berney}, \&
  {Sartori}}]{Schawinski2015}
{Schawinski}, K., {Koss}, M., {Berney}, S., \& {Sartori}, L.~F. 2015, \mnras,
  451, 2517, \dodoi{10.1093/mnras/stv1136}

\bibitem[{{Schindler} {et~al.}(2020){Schindler}, {Farina}, {Banados}, {Eilers},
  {Hennawi}, {Onoue}, {Venemans}, {Walter}, {Wang}, {Davies}, {Decarli}, {De
  Rosa}, {Drake}, {Fan}, {Mazzucchelli}, {Rix}, {Worseck}, \&
  {Yang}}]{Schindler2020}
{Schindler}, J.-T., {Farina}, E.~P., {Banados}, E., {et~al.} 2020, arXiv
  e-prints, arXiv:2010.06902.
\newblock \doarXiv{2010.06902}

\bibitem[{{Schmidt} {et~al.}(2019){Schmidt}, {Hennawi}, {Lee}, {Luki{\'c}},
  {O{\~n}orbe}, \& {White}}]{Schmidt2019}
{Schmidt}, T.~M., {Hennawi}, J.~F., {Lee}, K.-G., {et~al.} 2019, \apj, 882,
  165, \dodoi{10.3847/1538-4357/ab2fcb}

\bibitem[{{Schmidt} {et~al.}(2018){Schmidt}, {Hennawi}, {Worseck}, {Davies},
  {Luki{\'c}}, \& {O{\~n}orbe}}]{Schmidt2018}
{Schmidt}, T.~M., {Hennawi}, J.~F., {Worseck}, G., {et~al.} 2018, \apj, 861,
  122, \dodoi{10.3847/1538-4357/aac8e4}

\bibitem[{{Schmidt} {et~al.}(2017){Schmidt}, {Worseck}, {Hennawi}, {Prochaska},
  \& {Crighton}}]{Schmidt2017}
{Schmidt}, T.~M., {Worseck}, G., {Hennawi}, J.~F., {Prochaska}, J.~X., \&
  {Crighton}, N. H.~M. 2017, \apj, 847, 81, \dodoi{10.3847/1538-4357/aa83ac}

\bibitem[{{Shakura} \& {Sunyaev}(1973)}]{ShakuraSunyaev1973}
{Shakura}, N.~I., \& {Sunyaev}, R.~A. 1973, \aap, 500, 33

\bibitem[{{Shankar} {et~al.}(2010){Shankar}, {Weinberg}, \&
  {Shen}}]{Shankar2010b}
{Shankar}, F., {Weinberg}, D.~H., \& {Shen}, Y. 2010, \mnras, 406, 1959,
  \dodoi{10.1111/j.1365-2966.2010.16801.x}

\bibitem[{{Shen} {et~al.}(2007){Shen}, {Strauss}, {Oguri}, {Hennawi}, {Fan},
  {Richards}, {Hall}, {Gunn}, {Schneider}, {Szalay}, {Thakar}, {Vanden Berk},
  {Anderson}, {Bahcall}, {Connolly}, \& {Knapp}}]{Shen2007}
{Shen}, Y., {Strauss}, M.~A., {Oguri}, M., {et~al.} 2007, The Astronomical
  Journal, 133, 2222, \dodoi{10.1086/513517}

\bibitem[{{Shen} {et~al.}(2009){Shen}, {Strauss}, {Ross}, {Hall}, {Lin},
  {Richards}, {Schneider}, {Weinberg}, {Connolly}, {Fan}, {Hennawi}, {Shankar},
  {Vanden Berk}, {Bahcall}, \& {Brunner}}]{Shen2009}
{Shen}, Y., {Strauss}, M.~A., {Ross}, N.~P., {et~al.} 2009, \apj, 697, 1656,
  \dodoi{10.1088/0004-637X/697/2/1656}

\bibitem[{{Shull} {et~al.}(2004){Shull}, {Tumlinson}, {Giroux}, {Kriss}, \&
  {Reimers}}]{Shull2004}
{Shull}, J.~M., {Tumlinson}, J., {Giroux}, M.~L., {Kriss}, G.~A., \& {Reimers},
  D. 2004, \apj, 600, 570, \dodoi{10.1086/379924}

\bibitem[{{Simcoe} {et~al.}(2020){Simcoe}, {Onoue}, {Eilers}, {Banados},
  {Cooper}, {Furesz}, {Hennawi}, \& {Venemans}}]{Simcoe2020}
{Simcoe}, R.~A., {Onoue}, M., {Eilers}, A.-C., {et~al.} 2020, arXiv e-prints,
  arXiv:2011.10582.
\newblock \doarXiv{2011.10582}

\bibitem[{{Simcoe} {et~al.}(2012){Simcoe}, {Sullivan}, {Cooksey}, {Kao},
  {Matejek}, \& {Burgasser}}]{Simcoe2012}
{Simcoe}, R.~A., {Sullivan}, P.~W., {Cooksey}, K.~L., {et~al.} 2012, Nature,
  492, 79, \dodoi{10.1038/nature11612}

\bibitem[{{Springel} {et~al.}(2005){Springel}, {White}, {Jenkins}, {Frenk},
  {Yoshida}, {Gao}, {Navarro}, {Thacker}, {Croton}, {Helly}, {Peacock}, {Cole},
  {Thomas}, {Couchman}, {Evrard}, {Colberg}, \& {Pearce}}]{Springel2005}
{Springel}, V., {White}, S. D.~M., {Jenkins}, A., {et~al.} 2005, \nat, 435,
  629, \dodoi{10.1038/nature03597}

\bibitem[{{Stern} {et~al.}(2018){Stern}, {Faucher-Gigu{\`e}re}, {Hennawi},
  {Hafen}, {Johnson}, \& {Fielding}}]{Stern2018}
{Stern}, J., {Faucher-Gigu{\`e}re}, C.-A., {Hennawi}, J.~F., {et~al.} 2018,
  \apj, 865, 91, \dodoi{10.3847/1538-4357/aac884}

\bibitem[{{Stevans} {et~al.}(2014){Stevans}, {Shull}, {Danforth}, \&
  {Tilton}}]{Stevans2014}
{Stevans}, M.~L., {Shull}, J.~M., {Danforth}, C.~W., \& {Tilton}, E.~M. 2014,
  \apj, 794, 75, \dodoi{10.1088/0004-637X/794/1/75}

\bibitem[{{Suzuki}(2006)}]{Suzuki2006}
{Suzuki}, N. 2006, The Astrophysical Journal Supplement Series, 163, 110,
  \dodoi{10.1086/499272}

\bibitem[{{Taufik Andika} {et~al.}(2020){Taufik Andika}, {Jahnke}, {Onoue},
  {Ba{\~n}ados}, {Mazzucchelli}, {Novak}, {Eilers}, {Venemans}, {Schindler},
  {Walter}, {Neeleman}, {Simcoe}, {Decarli}, {Farina}, {Marian}, {Pensabene},
  {Cooper}, \& {Rojas}}]{Andika2020}
{Taufik Andika}, I., {Jahnke}, K., {Onoue}, M., {et~al.} 2020, arXiv e-prints,
  arXiv:2009.07784.
\newblock \doarXiv{2009.07784}

\bibitem[{{The Astropy Collaboration} {et~al.}(2018){The Astropy
  Collaboration}, {Price-Whelan}, {Sip{\H o}cz}, {G{\"u}nther}, {Lim},
  {Crawford}, {Conseil}, {Shupe}, {Craig}, {Dencheva}, {Ginsburg},
  {VanderPlas}, {Bradley}, {P{\'e}rez-Su{\'a}rez}, {de Val-Borro}, {Aldcroft},
  {Cruz}, {Robitaille}, {Tollerud}, {Ardelean}, {Babej}, {Bachetti}, {Bakanov},
  {Bamford}, {Barentsen}, {Barmby}, {Baumbach}, {Berry}, {Biscani}, {Boquien},
  {Bostroem}, {Bouma}, {Brammer}, {Bray}, {Breytenbach}, {Buddelmeijer},
  {Burke}, {Calderone}, {Cano Rodr{\'{\i}}guez}, {Cara}, {Cardoso},
  {Cheedella}, {Copin}, {Crichton}, {D{\'A}vella}, {Deil}, {Depagne},
  {Dietrich}, {Donath}, {Droettboom}, {Earl}, {Erben}, {Fabbro}, {Ferreira},
  {Finethy}, {Fox}, {Garrison}, {Gibbons}, {Goldstein}, {Gommers}, {Greco},
  {Greenfield}, {Groener}, {Grollier}, {Hagen}, {Hirst}, {Homeier}, {Horton},
  {Hosseinzadeh}, {Hu}, {Hunkeler}, {Ivezi{\'c}}, {Jain}, {Jenness}, {Kanarek},
  {Kendrew}, {Kern}, {Kerzendorf}, {Khvalko}, {King}, {Kirkby}, {Kulkarni},
  {Kumar}, {Lee}, {Lenz}, {Littlefair}, {Ma}, {Macleod}, {Mastropietro},
  {McCully}, {Montagnac}, {Morris}, {Mueller}, {Mumford}, {Muna}, {Murphy},
  {Nelson}, {Nguyen}, {Ninan}, {N{\"o}the}, {Ogaz}, {Oh}, {Parejko}, {Parley},
  {Pascual}, {Patil}, {Patil}, {Plunkett}, {Prochaska}, {Rastogi}, {Reddy
  Janga}, {Sabater}, {Sakurikar}, {Seifert}, {Sherbert}, {Sherwood-Taylor},
  {Shih}, {Sick}, {Silbiger}, {Singanamalla}, {Singer}, {Sladen}, {Sooley},
  {Sornarajah}, {Streicher}, {Teuben}, {Thomas}, {Tremblay}, {Turner},
  {Terr{\'o}n}, {van Kerkwijk}, {de la Vega}, {Watkins}, {Weaver}, {Whitmore},
  {Woillez}, \& {Zabalza}}]{astropy}
{The Astropy Collaboration}, {Price-Whelan}, A.~M., {Sip{\H o}cz}, B.~M.,
  {et~al.} 2018, ArXiv e-prints.
\newblock \doarXiv{1801.02634}

\bibitem[{{Tinker} {et~al.}(2010){Tinker}, {Robertson}, {Kravtsov}, {Klypin},
  {Warren}, {Yepes}, \& {Gottl{\"o}ber}}]{Tinker2010}
{Tinker}, J.~L., {Robertson}, B.~E., {Kravtsov}, A.~V., {et~al.} 2010, \apj,
  724, 878, \dodoi{10.1088/0004-637X/724/2/878}

\bibitem[{{Trainor} \& {Steidel}(2013)}]{TrainorSteidel2013}
{Trainor}, R., \& {Steidel}, C.~C. 2013, \apjl, 775, L3,
  \dodoi{10.1088/2041-8205/775/1/L3}

\bibitem[{{Tumlinson} {et~al.}(2011){Tumlinson}, {Thom}, {Werk}, {Prochaska},
  {Tripp}, {Weinberg}, {Peeples}, {O'Meara}, {Oppenheimer}, {Meiring}, {Katz},
  {Dav{\'e}}, {Ford}, \& {Sembach}}]{Tumlinson2011}
{Tumlinson}, J., {Thom}, C., {Werk}, J.~K., {et~al.} 2011, Science, 334, 948,
  \dodoi{10.1126/science.1209840}

\bibitem[{{van der Walt} {et~al.}(2011){van der Walt}, Colbert, \&
  Varoquaux}]{numpy}
{van der Walt}, S., Colbert, S.~C., \& Varoquaux, G. 2011, Computing in Science
  Engineering, 13, 22

\bibitem[{{Venemans} {et~al.}(2019){Venemans}, {Neeleman}, {Walter}, {Novak},
  {Decarli}, {Hennawi}, \& {Rix}}]{Venemans2019}
{Venemans}, B.~P., {Neeleman}, M., {Walter}, F., {et~al.} 2019, \apjl, 874,
  L30, \dodoi{10.3847/2041-8213/ab11cc}

\bibitem[{{Venemans} {et~al.}(2016){Venemans}, {Walter}, {Zschaechner},
  {Decarli}, {De Rosa}, {Findlay}, {McMahon}, \& {Sutherland}}]{Venemans2016}
{Venemans}, B.~P., {Walter}, F., {Zschaechner}, L., {et~al.} 2016, \apj, 816,
  37, \dodoi{10.3847/0004-637X/816/1/37}

\bibitem[{{Visbal} \& {Croft}(2008)}]{VisbalCroft2008}
{Visbal}, E., \& {Croft}, R. A.~C. 2008, \apj, 674, 660, \dodoi{10.1086/523843}

\bibitem[{{Vito} {et~al.}(2018){Vito}, {Brandt}, {Yang}, {Gilli}, {Luo},
  {Vignali}, {Xue}, {Comastri}, {Koekemoer}, {Lehmer}, {Liu}, {Paolillo},
  {Ranalli}, {Schneider}, {Shemmer}, {Volonteri}, \& {Wang}}]{Vito2018}
{Vito}, F., {Brandt}, W.~N., {Yang}, G., {et~al.} 2018, \mnras, 473, 2378,
  \dodoi{10.1093/mnras/stx2486}

\bibitem[{{Vito} {et~al.}(2019){Vito}, {Brandt}, {Bauer}, {Gilli}, {Luo},
  {Zamorani}, {Calura}, {Comastri}, {Mazzucchelli}, {Mignoli}, {Nanni},
  {Shemmer}, {Vignali}, {Brusa}, {Cappelluti}, {Civano}, \&
  {Volonteri}}]{Vito2019}
{Vito}, F., {Brandt}, W.~N., {Bauer}, F.~E., {et~al.} 2019, \aap, 628, L6,
  \dodoi{10.1051/0004-6361/201935924}

\bibitem[{{Volonteri}(2010)}]{Volonteri2010}
{Volonteri}, M. 2010, The Astronomy and Astrophysics Review, 18, 279,
  \dodoi{10.1007/s00159-010-0029-x}

\bibitem[{{Volonteri}(2012)}]{Volonteri2012}
---. 2012, Science, 337, 544, \dodoi{10.1126/science.1220843}

\bibitem[{{Walther} {et~al.}(2019){Walther}, {O{\~n}orbe}, {Hennawi}, \&
  {Luki{\'c}}}]{Walther2019}
{Walther}, M., {O{\~n}orbe}, J., {Hennawi}, J.~F., \& {Luki{\'c}}, Z. 2019,
  \apj, 872, 13, \dodoi{10.3847/1538-4357/aafad1}

\bibitem[{{Wang} {et~al.}(2020){Wang}, {Davies}, {Yang}, {Hennawi}, {Fan},
  {Barth}, {Jiang}, {Wu}, {Mudd}, {Ba{\~n}ados}, {Bian}, {Decarli}, {Eilers},
  {Farina}, {Venemans}, {Walter}, \& {Yue}}]{Wang2020}
{Wang}, F., {Davies}, F.~B., {Yang}, J., {et~al.} 2020, \apj, 896, 23,
  \dodoi{10.3847/1538-4357/ab8c45}

\bibitem[{{Wang} {et~al.}(2021){Wang}, {Yang}, {Fan}, {Hennawi}, {Barth},
  {Banados}, {Bian}, {Boutsia}, {Connor}, {Davies}, {Decarli}, {Eilers},
  {Farina}, {Green}, {Jiang}, {Li}, {Mazzucchelli}, {Nanni}, {Schindler},
  {Venemans}, {Walter}, {Wu}, \& {Yue}}]{Wang2021}
{Wang}, F., {Yang}, J., {Fan}, X., {et~al.} 2021, \apjl, 907, L1,
  \dodoi{10.3847/2041-8213/abd8c6}

\bibitem[{{White} {et~al.}(2008){White}, {Martini}, \& {Cohn}}]{White2008}
{White}, M., {Martini}, P., \& {Cohn}, J.~D. 2008, \mnras, 390, 1179,
  \dodoi{10.1111/j.1365-2966.2008.13817.x}

\bibitem[{{White} {et~al.}(2012){White}, {Myers}, {Ross}, {Schlegel},
  {Hennawi}, {Shen}, {McGreer}, {Strauss}, {Bolton}, {Bovy}, {Fan},
  {Miralda-Escude}, {Palanque-Delabrouille}, {Paris}, {Petitjean}, {Schneider},
  {Viel}, {Weinberg}, {Yeche}, {Zehavi}, {Pan}, {Snedden}, {Bizyaev},
  {Brewington}, {Brinkmann}, {Malanushenko}, {Malanushenko}, {Oravetz},
  {Simmons}, {Sheldon}, \& {Weaver}}]{White2012}
{White}, M., {Myers}, A.~D., {Ross}, N.~P., {et~al.} 2012, \mnras, 424, 933,
  \dodoi{10.1111/j.1365-2966.2012.21251.x}

\bibitem[{{Worseck} {et~al.}(2021){Worseck}, {Khrykin}, {Hennawi}, {Prochaska},
  \& {Farina}}]{Worseck2021}
{Worseck}, G., {Khrykin}, I.~S., {Hennawi}, J.~F., {Prochaska}, J.~X., \&
  {Farina}, E.~P. 2021, arXiv e-prints, arXiv:2101.01196.
\newblock \doarXiv{2101.01196}

\bibitem[{{Worseck} {et~al.}(2014){Worseck}, {Prochaska}, {O'Meara}, {Becker},
  {Ellison}, {Lopez}, {Meiksin}, {M{\'e}nard}, {Murphy}, \&
  {Fumagalli}}]{Worseck2014}
{Worseck}, G., {Prochaska}, J.~X., {O'Meara}, J.~M., {et~al.} 2014, \mnras,
  445, 1745, \dodoi{10.1093/mnras/stu1827}

\bibitem[{{Wyithe} \& {Bolton}(2011)}]{WyitheBolton2011}
{Wyithe}, J.~S.~B., \& {Bolton}, J.~S. 2011, Monthly Notices of the Royal
  Astronomical Society, 412, 1926, \dodoi{10.1111/j.1365-2966.2010.18030.x}

\bibitem[{{Yang} {et~al.}(2020{\natexlab{a}}){Yang}, {Wang}, {Fan}, {Hennawi},
  {Davies}, {Yue}, {Eilers}, {Farina}, {Wu}, {Bian}, {Pacucci}, \&
  {Lee}}]{Yang2020b}
{Yang}, J., {Wang}, F., {Fan}, X., {et~al.} 2020{\natexlab{a}}, \apj, 904, 26,
  \dodoi{10.3847/1538-4357/abbc1b}

\bibitem[{{Yang} {et~al.}(2020{\natexlab{b}}){Yang}, {Wang}, {Fan}, {Hennawi},
  {Davies}, {Yue}, {Banados}, {Wu}, {Venemans}, {Barth}, {Bian}, {Boutsia},
  {Decarli}, {Farina}, {Green}, {Jiang}, {Li}, {Mazzucchelli}, \&
  {Walter}}]{Yang2020a}
---. 2020{\natexlab{b}}, \apjl, 897, L14, \dodoi{10.3847/2041-8213/ab9c26}

\bibitem[{{Yu} \& {Tremaine}(2002)}]{YuTremaine2002}
{Yu}, Q., \& {Tremaine}, S. 2002, \mnras, 335, 965,
  \dodoi{10.1046/j.1365-8711.2002.05532.x}

\end{thebibliography}

\end{document}